\documentclass[letterpaper,ngerman,english]{article}
\usepackage[T1]{fontenc}
\usepackage[latin9]{inputenc}
\usepackage{float}
\usepackage{textcomp}
\usepackage{amsmath}
\usepackage{amssymb}
\usepackage{graphicx}

\makeatletter

\providecommand{\tabularnewline}{\\}

\newcommand{\lyxaddress}[1]{
\par {\raggedright #1
\vspace{1.4em}
\noindent\par}
}

\makeatother

\usepackage{babel}
\begin{document}
\date{}

\title{Decawave UWB clock drift correction and power self-calibration}

\author{Juri Sidorenko \textsuperscript{ab*}, Volker Schatz \textsuperscript{a},\\
Norbert Scherer-Negenborn \textsuperscript{a}, Michael Arens \textsuperscript{a},
\\
Urs Hugentobler \textsuperscript{b}}
\maketitle

\lyxaddress{\textsuperscript{a}Fraunhofer Institute of Optronics, System Technologies
and Image Exploitation IOSB, Germany - juri.sidorenko@iosb.fraunhofer.de\\
\textsuperscript{b}Institute of Astronomical and Physical Geodesy,
Technical University of Munich, Germany - urs.hugentobler@tum.de}


\section{Abstract}

The position accuracy based on Decawave Ultra-Wideband (UWB) is affected
mainly by three factors: hardware delays, clock drift, and signal
power. This article discusses the last two factors. The general approach
to clock drift correction uses the phase-locked loop (PLL) integrator,
which we show is subject to signal power variations, and therefore,
is less suitable for clock drift correction. The general approach
to the estimation of signal power correction curves requires additional
measurement equipment. This article presents a new method for obtaining
the curve without additional hardware and clock drift correction without
the PLL integrator. Both correction methods were fused together to
improve two-way ranging (TWR). 

\section{Introduction}

In the last century, autonomous systems became omnipresent in almost
every field of the industry. Spending on robotics is expected to reach
67 billion US dollars by 2025, as compared to 11 billion in 2005 \cite{RobotMarket}.
One of the most important tasks in robotics is the interaction between
a robot and its environment. This task can only be accomplished if
the location of the robot with respect to its environment is known.
Visual sensors are very common for localization \cite{SLAM1,SLAM2}.
In some cases, estimating the position in non-line-of-sight conditions
is required. Radio-frequency-based (RF) sensors are able to operate
in such conditions, but the outcome depends highly on measurement
principles, such as received signal strength indicator (RSSI) \cite{RSSI},
fingerprinting \cite{fingerPrinting}, FMCW \cite{LPM} and UWB \cite{UWB_Limits},
as well as on techniques such as the angle of arrival \cite{angleOfarrival},
time of arrival \cite{TWR} or time difference of arrival \cite{TDOA}.
Indoor positioning is, in general, a challenge for RF-based localization
systems. Reflections could cause interference with the main signal.
In contrast to narrowband signals are ultra-wideband (UWB) signals,
which are more robust to fading \cite{fading1,fading2}. A common
UWB system is the Decawave UWB transceiver \cite{Why_Decawave}, which
is low cost and provides centimeter precision. The accuracy and precision
of this chip are affected by three factors: hardware delays, clock
drift, and signal power \cite{Decawave_clock,Signal_power}. This
article discusses clock drift correction and signal power error, which
is specific to the Decawave UWB transceiver and affects the accuracy
of the position significantly. The general approach to estimating
signal power dependency is to use ground truth data, which are provided
by additional measurement equipment \cite{SignalPower_correction}.
The clock drift error is caused by the different frequencies of the
transceiver clocks. The general approach to Decawave UWB clock drift
correction is to use the integrator of the phase-locked loop (PLL)
\cite{PLL_1,PLL_2,Cico,PLL_Decawave2018}. In the following section,
we explain that the general approach to clock drift correction is
not suitable because the PLL is also affected by the signal power.
Therefore, a more accurate method for clock drift correction is presented.
The middle sections of this article discuss the estimation of the
signal power correction curve without the need for additional hardware.
As far as we know, nobody has obtained a signal power correction curve
by self-calibration before. The last part of this article presents
a two-way ranging (TWR) method that is able to use the correction
methods for distance estimation. 

\begin{table}[H]
\begin{centering}
\caption{Notations used\label{tab:notations}}
\ \\
\par\end{centering}
\centering{}%
\begin{tabular}{|c|c|}
\hline 
Notations & \selectlanguage{ngerman}%
Definition\selectlanguage{english}%
\tabularnewline
\hline 
\hline 
$T_{i}$ & Timestamp\tabularnewline
\hline 
$\Delta T_{n,m}$ & Difference between two timestamps $T_{m}-T_{n}$\tabularnewline
\hline 
$C_{n,m}$ & Clock drift with respect to the timestamps n and m\tabularnewline
\hline 
$E_{i}$ & Timestamp error due to signal power\tabularnewline
\hline 
$Z$ & Hardware delay and signal power correction offset\tabularnewline
\hline 
\end{tabular}
\end{table}

\section{Decawave UWB}

Decawave transceivers are based on UWB technology and are compliant
with IEEE802.15.4-2011 standards \cite{Decawave_Anaysis}. They support
six frequency bands with center frequencies from 3.5 GHz to 6.5 GHz
and data rates of up to 6.8 Mb/s. The bandwidth varies with the selected
center frequencies from 500 up to 1,000 MHz. With higher bandwidth,
the send impulse becomes sharper. The timestamps for the positioning
are provided by an estimation of the channel impulse response, which
is obtained by correlating a known preamble sequence against the received
signal and accumulating the result over a period of time. In contrast
to narrowband signals is UWB more resistant to multipath fading. Reflections
would cause an additional peak in the impulse response. The probability
that two peaks interfere with each other is small. The sampling of
the signal is performed by an internal 64 GHz chip with 15 ps event-timing
precision (4.496 mm). Because of general regulations, the transmit
power density is limited to \textminus 41.3 dBm/MHz. These regulations
are due to the high bandwidth occupied by the UWB transceiver. The
maximum permissible power level is averaged over 1 ms period; hence,
the power can be increased for shorter message durations. The following
experiments were carried out with the Decawave EVK1000. This board
mainly consists of a DW1000 chip and an STM32 ARM processor. 

\section{Clock drift correction}

In practice, it is not possible to manufacture exactly the same clock
generators, so every transceiver has a different clock frequency.
Clock drift correction represents the difference between clock frequencies
but not current time values.

\subsection{General approach}

\begin{figure}[H]
\begin{centering}
\includegraphics[scale=0.5]{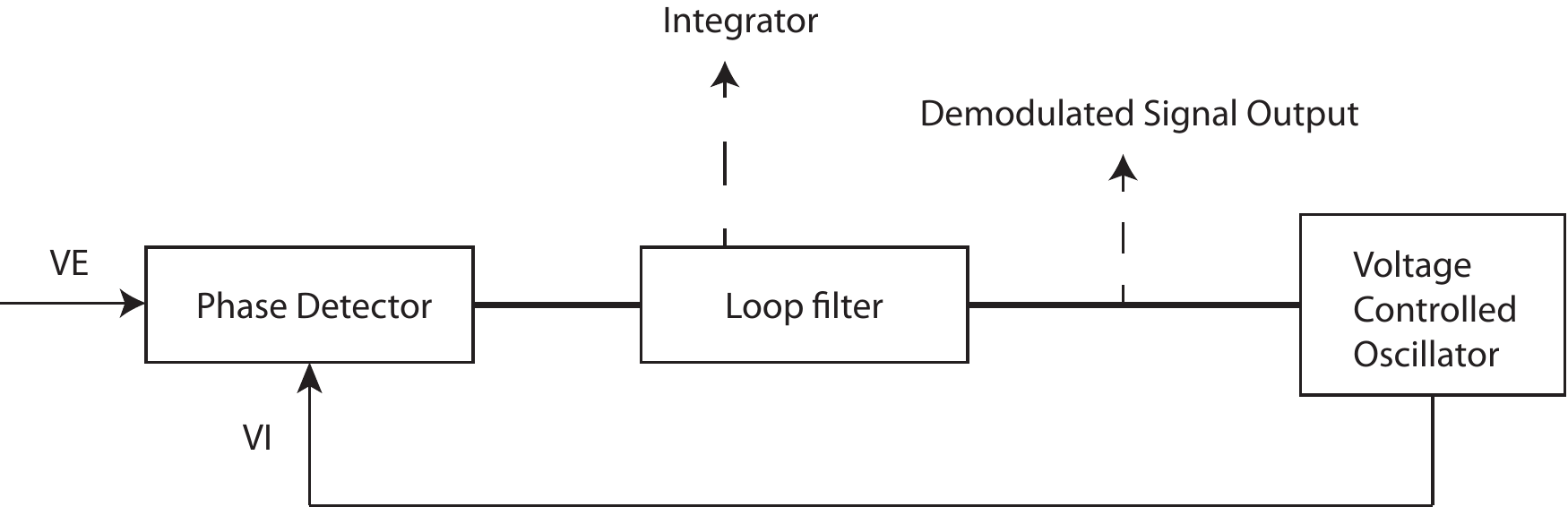}
\par\end{centering}
\caption{Example of the phase locked loop (PLL)\label{fig:PLL}}
\end{figure}

The general approach to clock drift correction is to use the PLL integrator
\cite{PLL_1,PLL_2,Cico,PLL_Decawave2018}. Figure \ref{fig:PLL} shows
an example of frequency demodulation by a PLL. The voltage-controlled
oscillator (VCO) is set to the mid-position and the loop is locked
in at the frequency of the carrier wave. Modulations on the carrier
would cause the VCO frequency to follow the incoming signal, so changes
in the voltage correspond to the applied modulations. The difference
between the received carrier frequency (VE) and the internal loop
frequency (VI) can be observed in the integrator of the loop filter. 

\begin{figure}[H]
\begin{centering}
\includegraphics[scale=0.6]{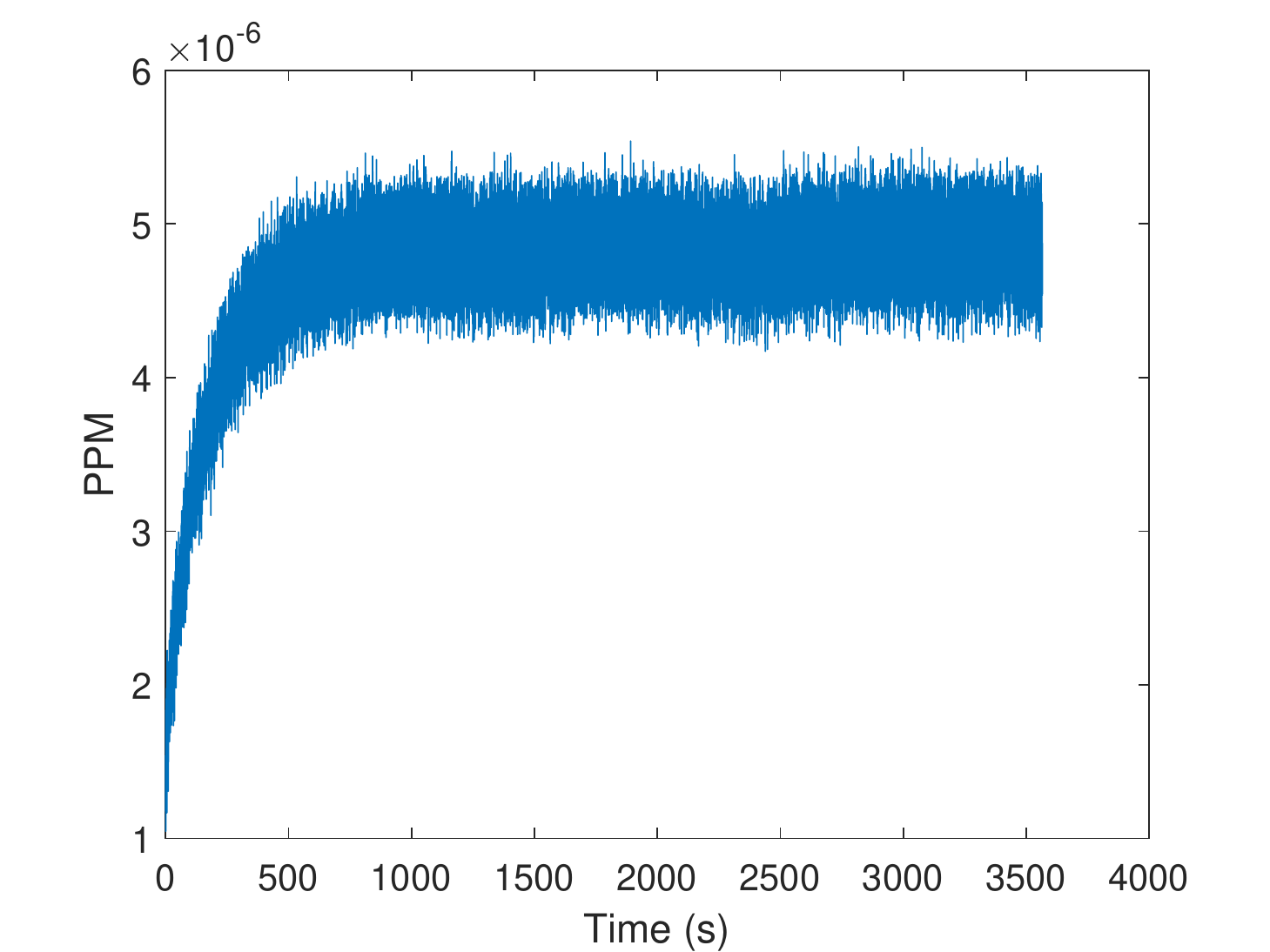}
\par\end{centering}
\caption{Integrator of the PLL\label{fig:IntegratorPLL}}
\end{figure}

In Figure \ref{fig:IntegratorPLL}, the integrator output is presented.
The test scenario is based on measurements obtained at every 50 ms
between two stationary transceivers. The difference between the two
frequencies is about five parts per million. Reaching the final condition
took up to 15 minutes.

\begin{figure}[H]
\begin{centering}
\includegraphics[scale=0.6]{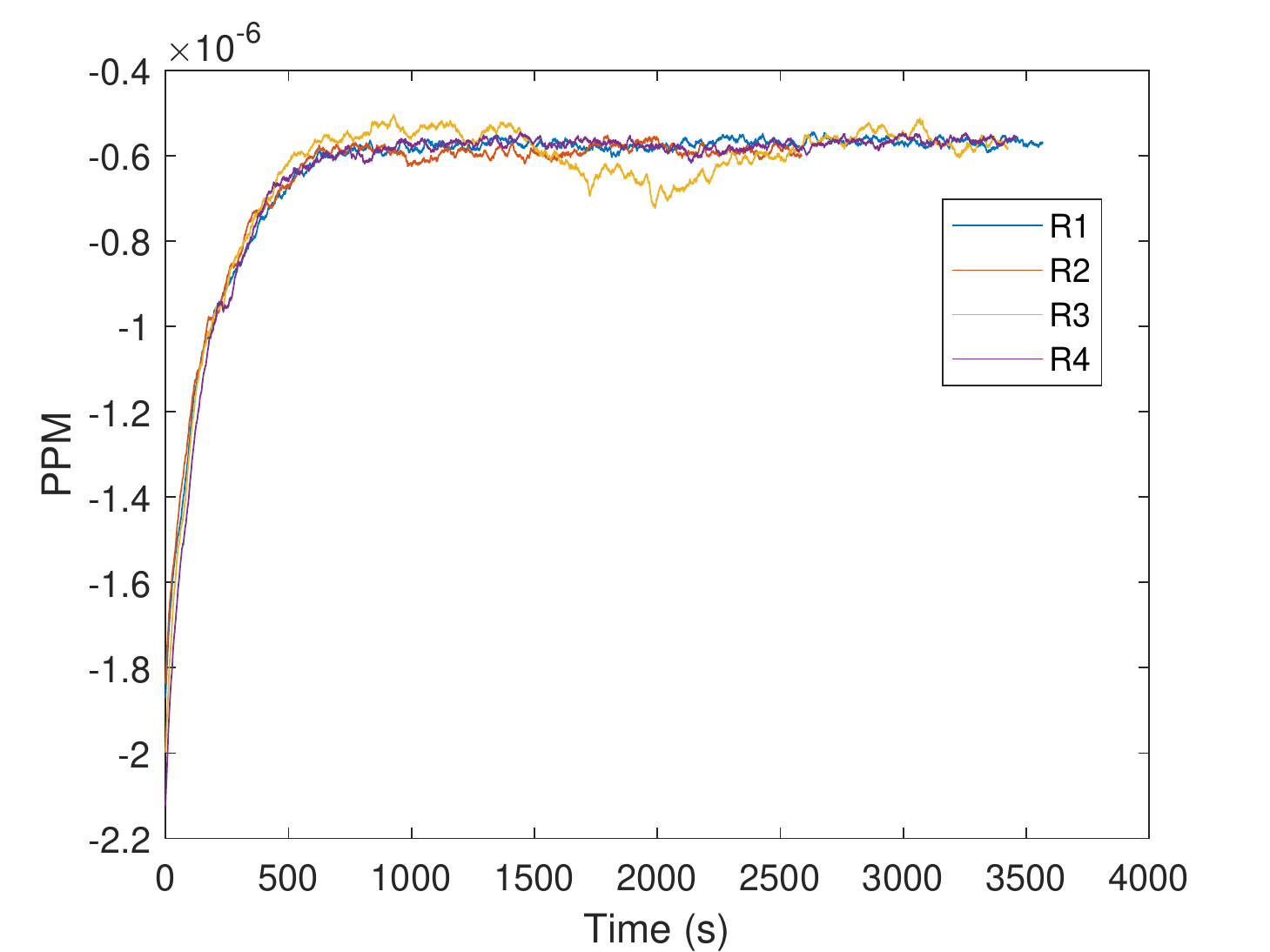}
\par\end{centering}
\caption{Filtered integrator of the PLL four times restarted\label{fig:IntegratorRestart}}
\end{figure}

The tests were repeated four times with another two stationary stations.
Figure \ref{fig:IntegratorRestart} shows the filtered results of
the obtained curves provided by a 500-point moving average filter.
The curve progression is deterministic.

\begin{figure}[H]
\begin{centering}
\includegraphics[scale=0.4]{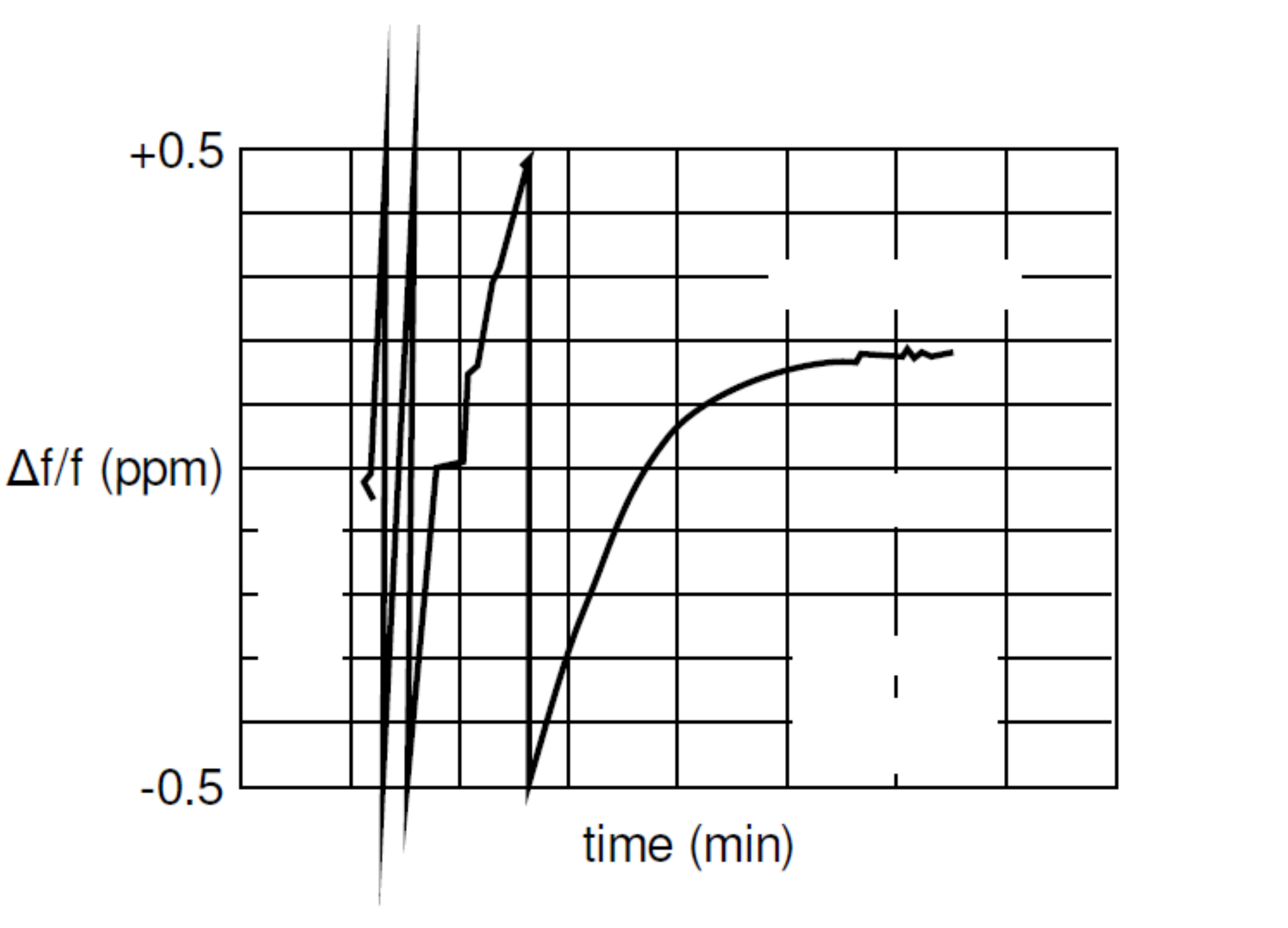}
\par\end{centering}
\caption{DW1000 temperature crystal oscillator warm-up\label{fig:DW1000-temperature-crystal}
\cite{SignalPower_correction}, used with permission}
\end{figure}

Decawave indicates that the logarithmic increase of the integrator
at the beginning is due to the warm-up when the crystal oscillator
is activated, graphically represented in Figure \ref{fig:DW1000-temperature-crystal}.
This oscillator follows from the combination of a quartz crystal and
the circuitry within the DW1000-based design.

\begin{figure}[H]
\begin{centering}
\includegraphics[scale=0.3]{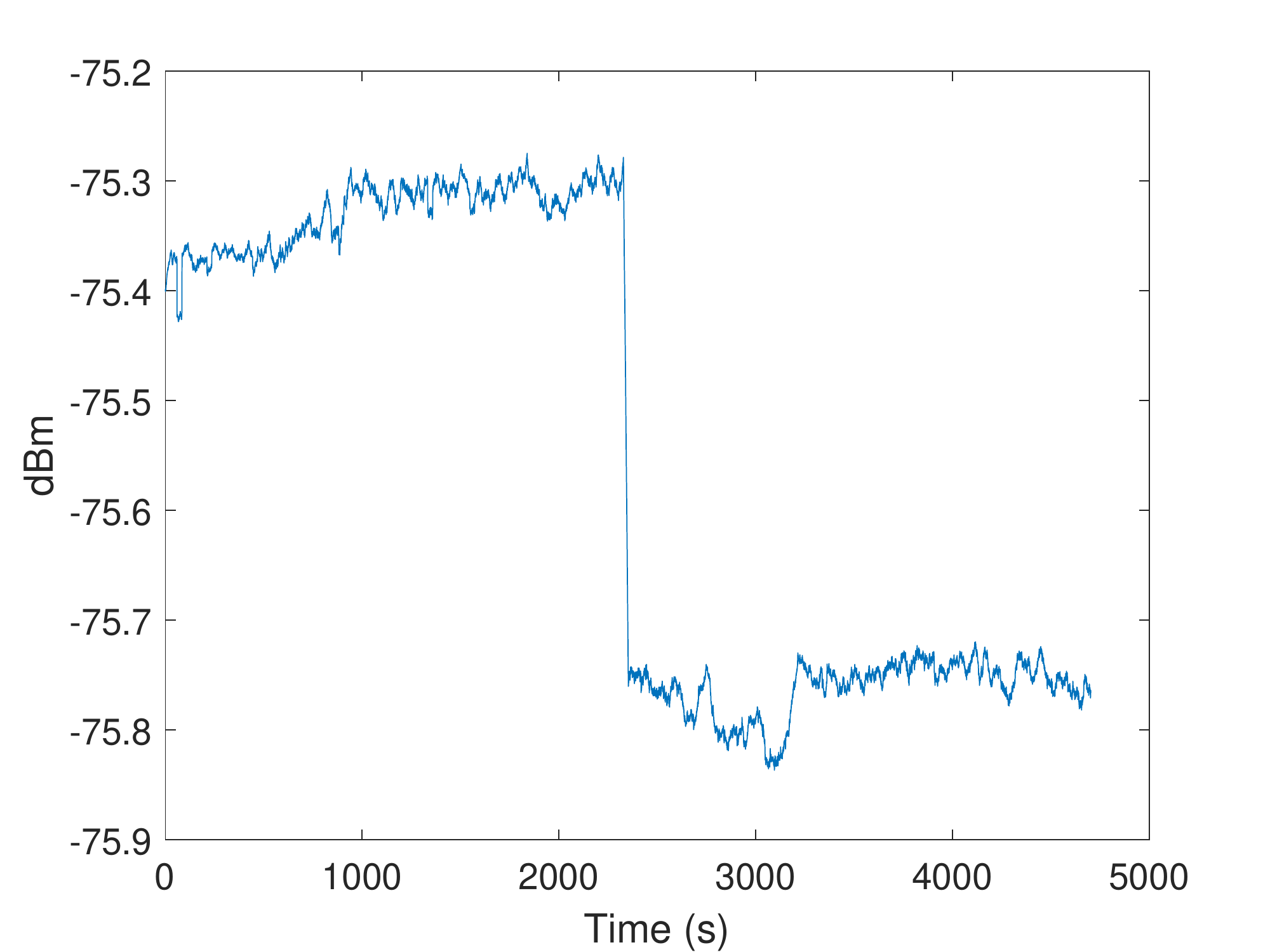}\includegraphics[scale=0.3]{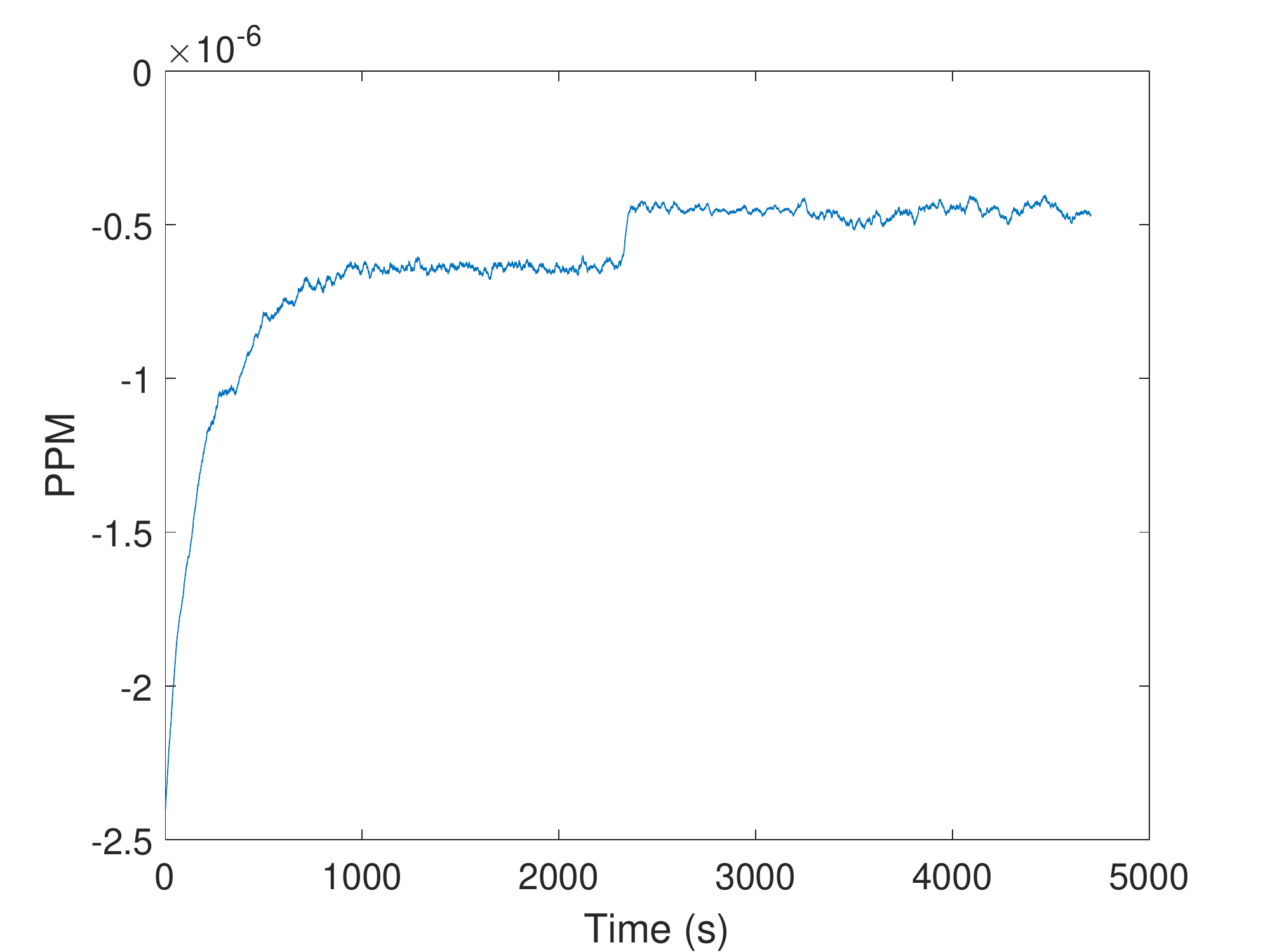}
\par\end{centering}
\caption{Left: Filtered received signal power. Right: Filtered integrator of
the PLL \label{fig:Left:-Power_and_Integrator} }
\end{figure}

In the following test scenario, the effect of the signal power on
the integrator was investigated. Both the transmitter and receiver
stations were stationary. The left side of Figure \ref{fig:Left:-Power_and_Integrator}
shows the measured signal strength at the receiving station. After
about 4,600 measurements, we arranged the transmitter to reduce the
signal power. The integrator of the receiver jumped after the signal
power changed to a new level, indicating that distance changes between
the transmitter and receiver would affect the integrator, and so,
affect the clock drift correction as well. 

The reason for this dependency could be the analog phase detectors
of the PLL, in which the loop gain$K_{D}$ is a function of amplitude,
which affects the error signal $v_{e}(t)=K_{D}[\varPhi_{Out}(t)-\varPhi_{In}(t)]$
, and so, affects the pull-in time (total time taken by the PLL to
lock) as well.

\subsection{Proposed approach for the clock drift correction}

\begin{figure}[H]
\begin{centering}
\includegraphics[scale=0.4]{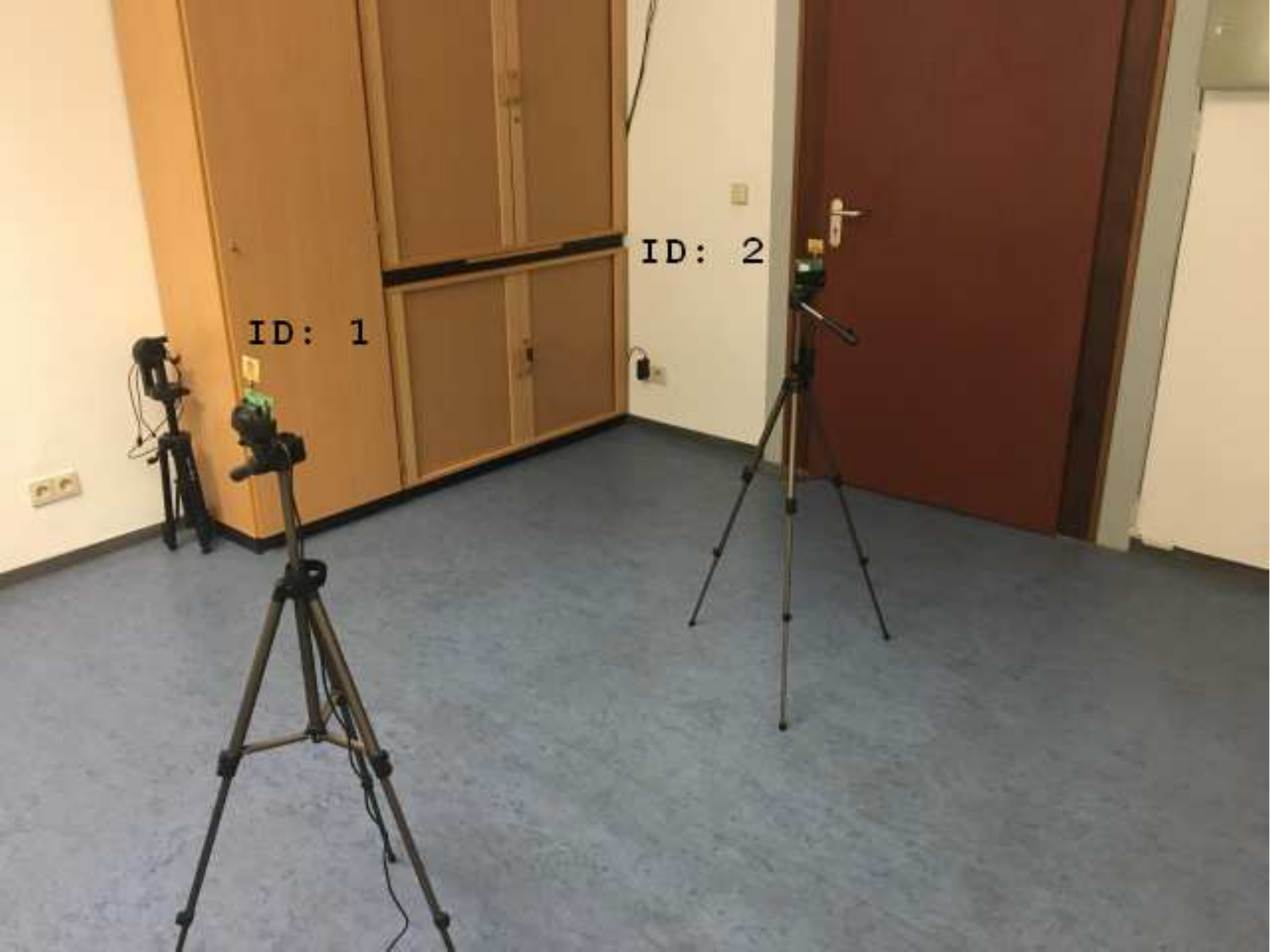}
\par\end{centering}
\caption{Measurement setup \label{fig:Measurement-setup}}
\end{figure}

In this section, we present an alternative method for the clock drift
correction, which is independent of the signal power. The measurement
setup is presented in Figure \ref{fig:Measurement-setup}. All measurements
and calibrations were conducted with Decawave EVK1000 boards. The
station with the identification number (id) 2 is the transmitting
station (TX). The receiving station (RX) has the identification number
1. The receiving signal power, as well as the timestamps, were obtained
by reading the register provided by the transceivers \cite{Signal_power,SignalPower_correction}.

The general settings for the hardware setup can be found in Table
\ref{tab:Test-settings} and the notations in Table \ref{tab:notations}.

\begin{table}[H]
\begin{centering}
\begin{tabular}{|c|c|}
\hline 
Channel & 2\tabularnewline
\hline 
Center Frequency & 3993.6 MHz\tabularnewline
\hline 
Bandwidth & 499.2 MHz\tabularnewline
\hline 
Pulse repetition frequency & 64 MHz\tabularnewline
\hline 
Preamble length & 128\tabularnewline
\hline 
Data rate & 6.81 Mbps\tabularnewline
\hline 
\end{tabular}
\par\end{centering}
\caption{Test settings \label{tab:Test-settings}}
\end{table}

\begin{figure}[H]
\begin{centering}
\includegraphics[scale=0.4]{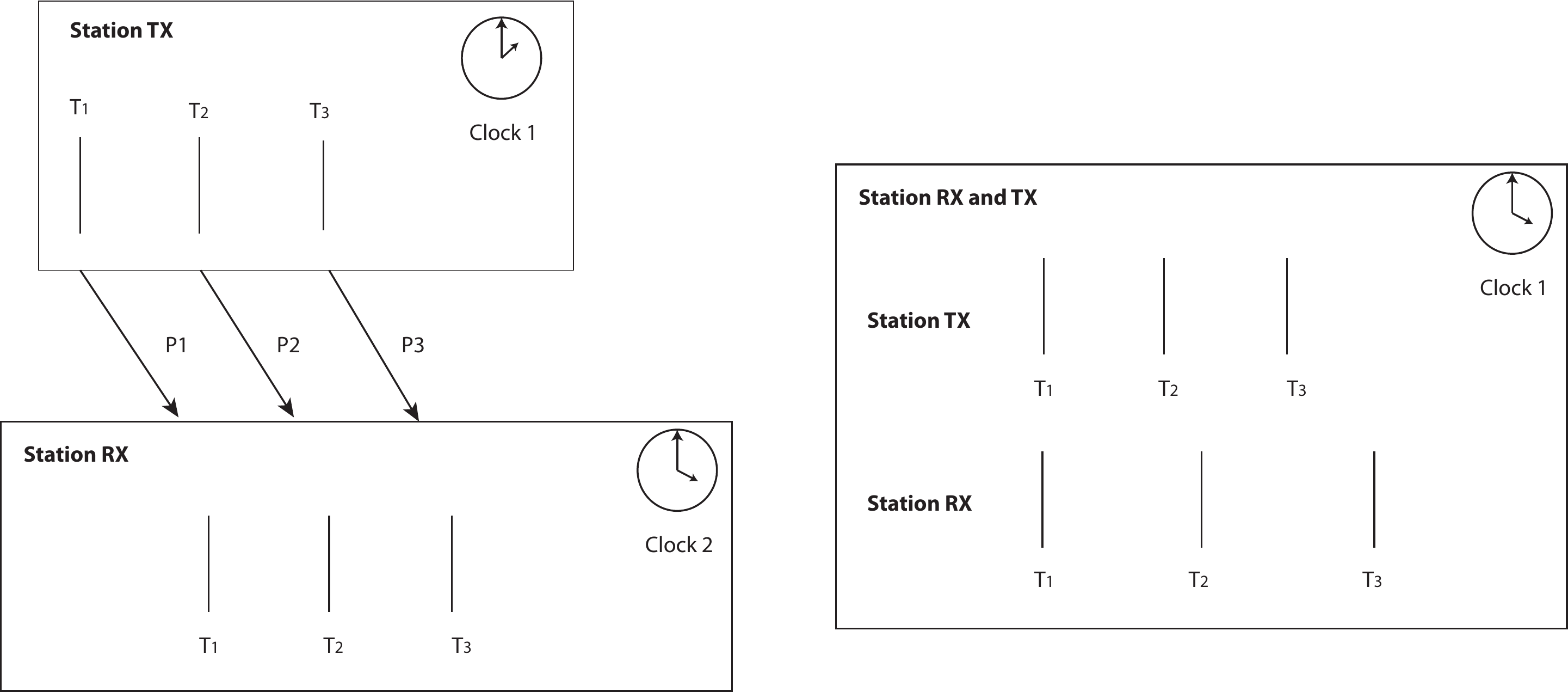}
\par\end{centering}
\caption{Alternative clock drift correction \label{fig:Alternative-clock-drift}}
\end{figure}

Figure \ref{fig:Alternative-clock-drift} shows a schematic diagram
of the approach. TX is sending three signals at times $T_{1}$, $T_{2}$,
and $T_{3}$. The clocks of the transmitter and receiver are not synchronous.
If the clocks have no drift, then both clocks should have the same
frequency and the difference between $\Delta T_{1,2}=T_{2}-T_{1}$
should be the same for the transmitter and the receiver; otherwise,
$\Delta T_{1,2}^{RX}\neq\Delta T_{1,2}^{TX}$ . The same applies to
$\Delta T_{1,3}$. If the clock of RX is running faster than that
of TX, then $\Delta T_{1,3}^{RX}>\Delta T_{1,3}^{TX}$ and the clock
drift error becomes $C_{1,2}=\Delta T_{1,2}^{RX}-\Delta T_{1,2}^{TX}$
.

Previously, the frequency difference between the two clocks was presented
by the integrator of the PLL. After the warm-up time, the clocks reached
their final frequencies. The clock error now increased linearly. For
short measurement periods the clock drift error can be assumed to
be linear even during the the oscillator\textquoteright s warm-up.

\begin{figure}[H]
\begin{centering}
\includegraphics[scale=0.4]{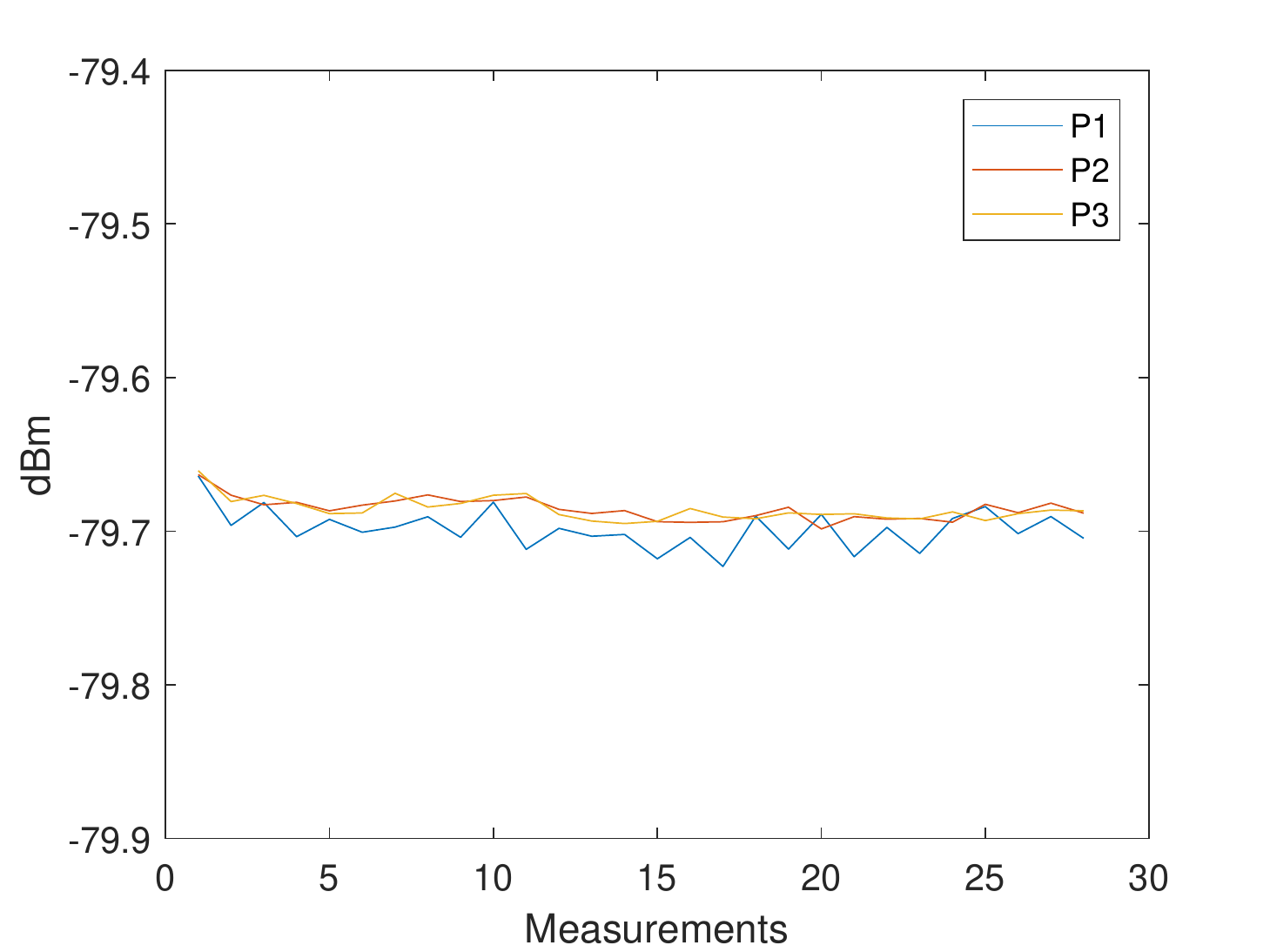}\includegraphics[scale=0.4]{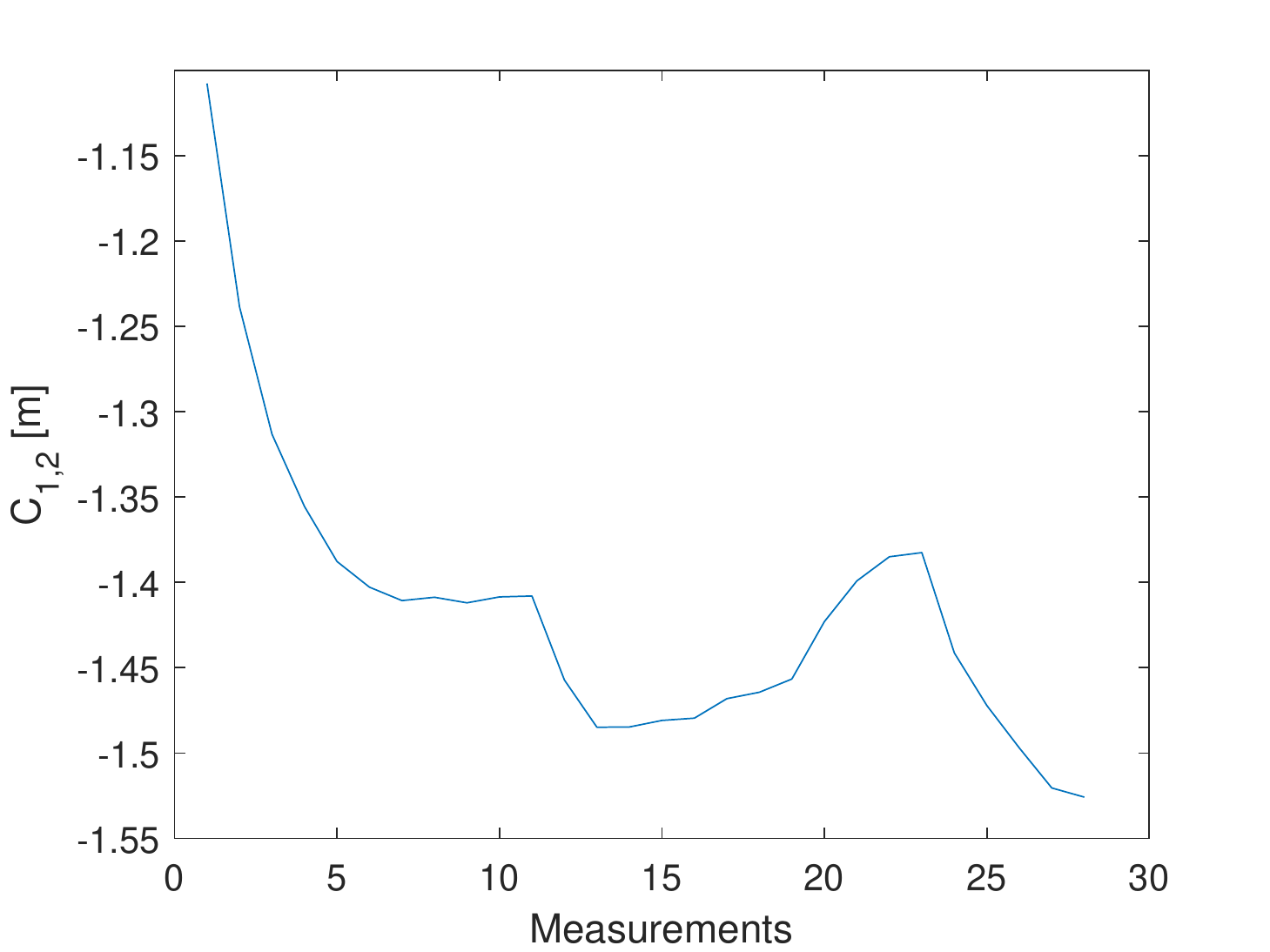}
\par\end{centering}
\caption{Left: Signal strength. Right: Error due to clock drift\label{fig:Clock_drift_error}}
\end{figure}

The main idea is for the clock drift error $C_{1,3}=\Delta T_{1,3}^{RX}-\Delta T_{1,3}^{TX}$
to be used for correcting the timestamp $T_{2}$ by simple linear
interpolation. In Figure \ref{fig:Clock_drift_error}, three messages,
P1, P2, and P3, with constant signal powers have been sent. The delay
between every message was about 2 ms. The values are already filtered;
hence, every point consists of the mean of 4,000 measurements. The
right side of Figure \ref{fig:Clock_drift_error} shows the clock
drift error $C_{1,2}=\Delta T_{1,2}^{RX}-\Delta T_{1,2}^{TX}$ . Because
of the long delay is the distance error resulting from the clock drift
about 1 m. 

In the next step, the clock drift error $C_{1,2}$ is corrected by
the linear interpolation of $C_{1,3}$. 

\begin{equation}
C_{1,2}^{'}=C_{1,2}-\frac{C_{1,3}}{\Delta T_{1,3}^{TX}}\cdotp\Delta T_{1,2}^{TX}\label{eq:Correction}
\end{equation}

\begin{figure}[H]
\begin{centering}
\includegraphics[scale=0.4]{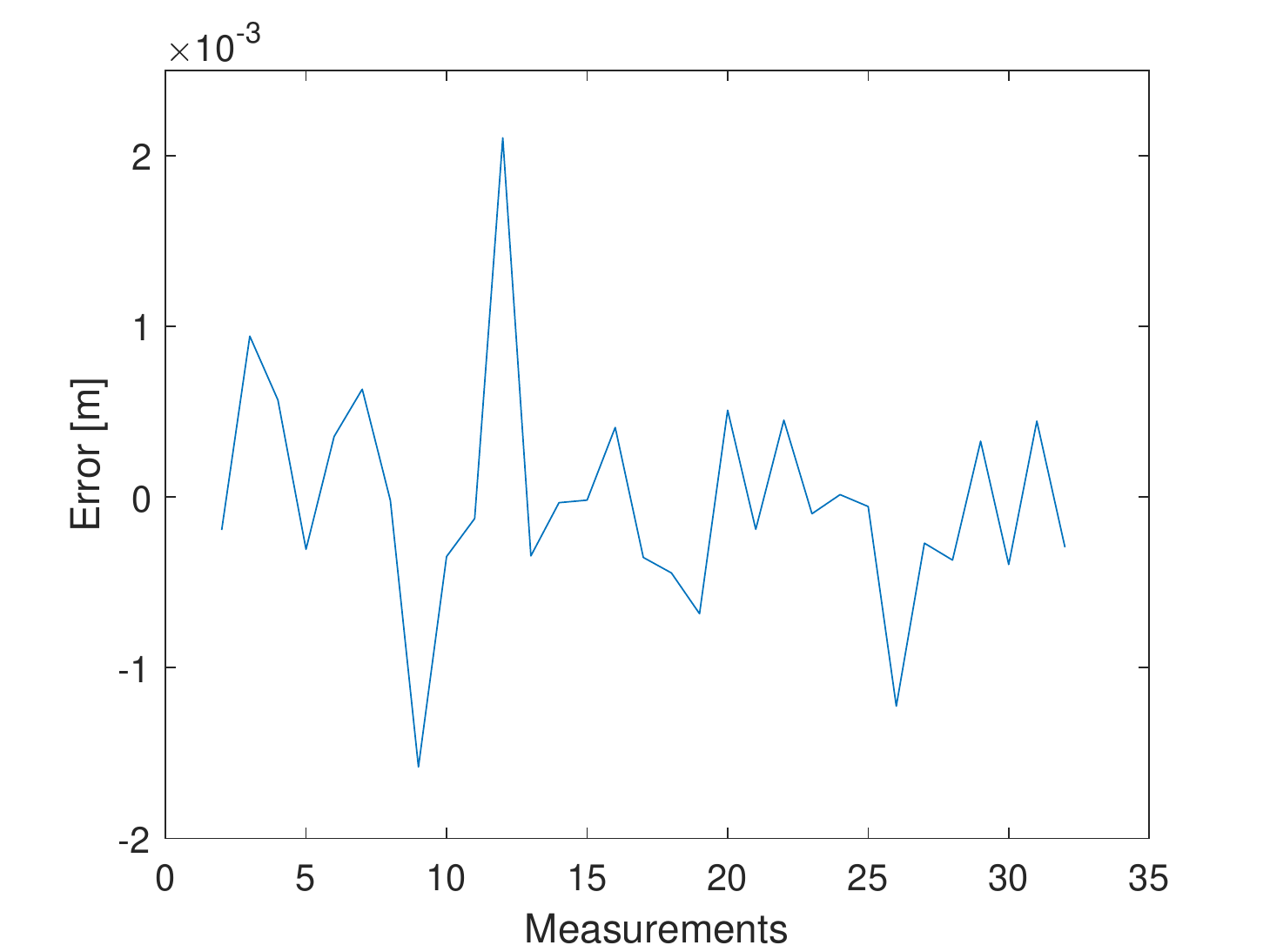}
\par\end{centering}
\caption{Results of clock drift correction $C_{1,2}^{'}$ \label{fig:Clock-drift-correction}}
\end{figure}

The results are shown in Figure \ref{fig:Clock-drift-correction}.
The correction requires only three messages and the remaining average
offset is about $-1.915\cdotp10^{-5}m$. The linear interpolation
is also suitable for the warm-up phase. The implementation of the
presented clock drift correction for the TWR is presented in section
\ref{sec:Two-way-ranging}. A position error caused by a constant
velocity of the object is also corrected by the linear interpolation,
due to the linear increase of the position error (pseudo clock drift).
In pratise, is it possible to obtain $\Delta T_{1,3}^{TX}\thickapprox1\,ms$.
An acceleration high enough to cause an error greater than 5 mm, would
require near most 1,000g $\left(10^{4}\frac{m}{s^{2}}\right)$.

\section{Signal power correction}

The next section discusses the signal power correction. It is known
that the time stamp of the DW1000 is affected by the signal power,
in which an increase causes a negative shift of the time stamp and
vice versa. 

\subsection{General approach}

\begin{figure}[H]
\begin{centering}
\includegraphics[scale=0.35]{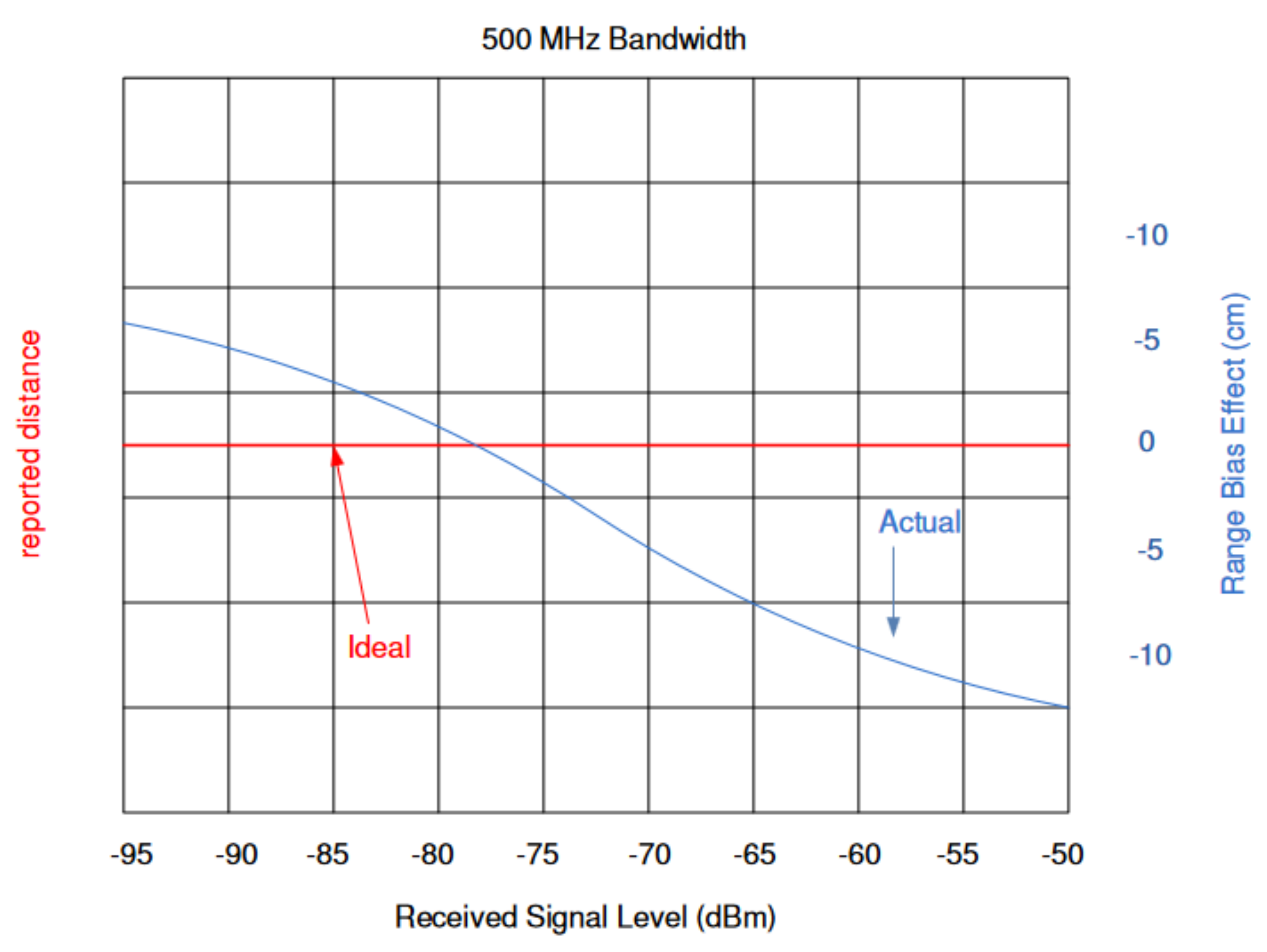}
\par\end{centering}
\caption{Effect of range bias on reported distance \cite{SignalPower_correction}\label{fig:Sigmoid},
used with permission}
\end{figure}

Figure \ref{fig:Sigmoid} illustrates the reported distance error
with respect to the received signal power. At a certain signal strength,
the range bias effect should be zero. In Figure \ref{fig:Sigmoid}
the bias vanishes between  \textminus 80 and \textminus 75 dBm. The
correction curve is affected by the system design elements, such as
printed circuit boards, antenna gain, and pulse repetition frequency
(PRF). The general approach to correction curve estimation is to compare
the distance measurements with the ground truth distances. This method
has two disadvantages. First of all, additional measurement equipment
is necessary. Second, every created curve applies to two stations
but not every individual station. 

\begin{figure}[H]
\begin{centering}
\includegraphics[scale=0.35]{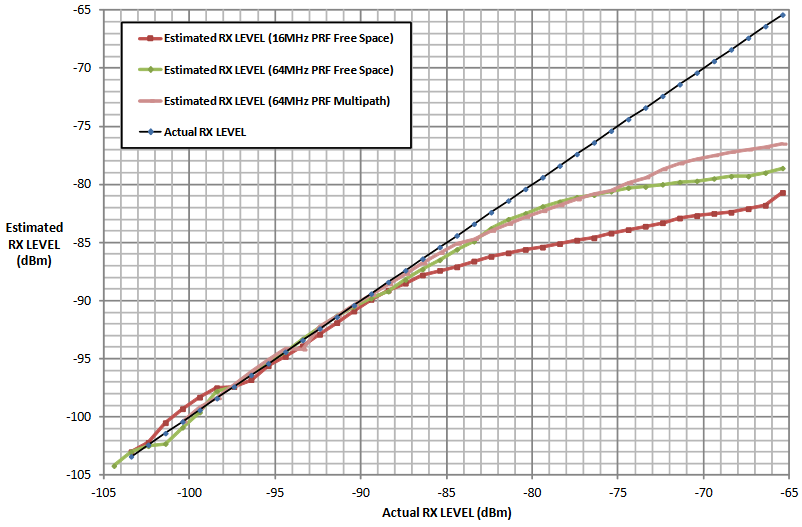}
\par\end{centering}
\caption{Estimated RX level with respect to actual RX level \cite{Signal_power}\label{fig:Estimated-RX-level},
used with permission}
\end{figure}

Figure \ref{fig:Estimated-RX-level} shows the relationship between
the measured and correct signal strengths for different PRF. The measured
signal power is correct only for measurements smaller than \textminus 85
dBm. The knowledge of the difference between the measured and correct
signal strengths can be used for additional measurement techniques,
such as the RSSI, for distance estimation.

\subsection{Proposed approach for the signal power correction}

\begin{figure}[H]
\begin{centering}
\includegraphics[scale=0.4]{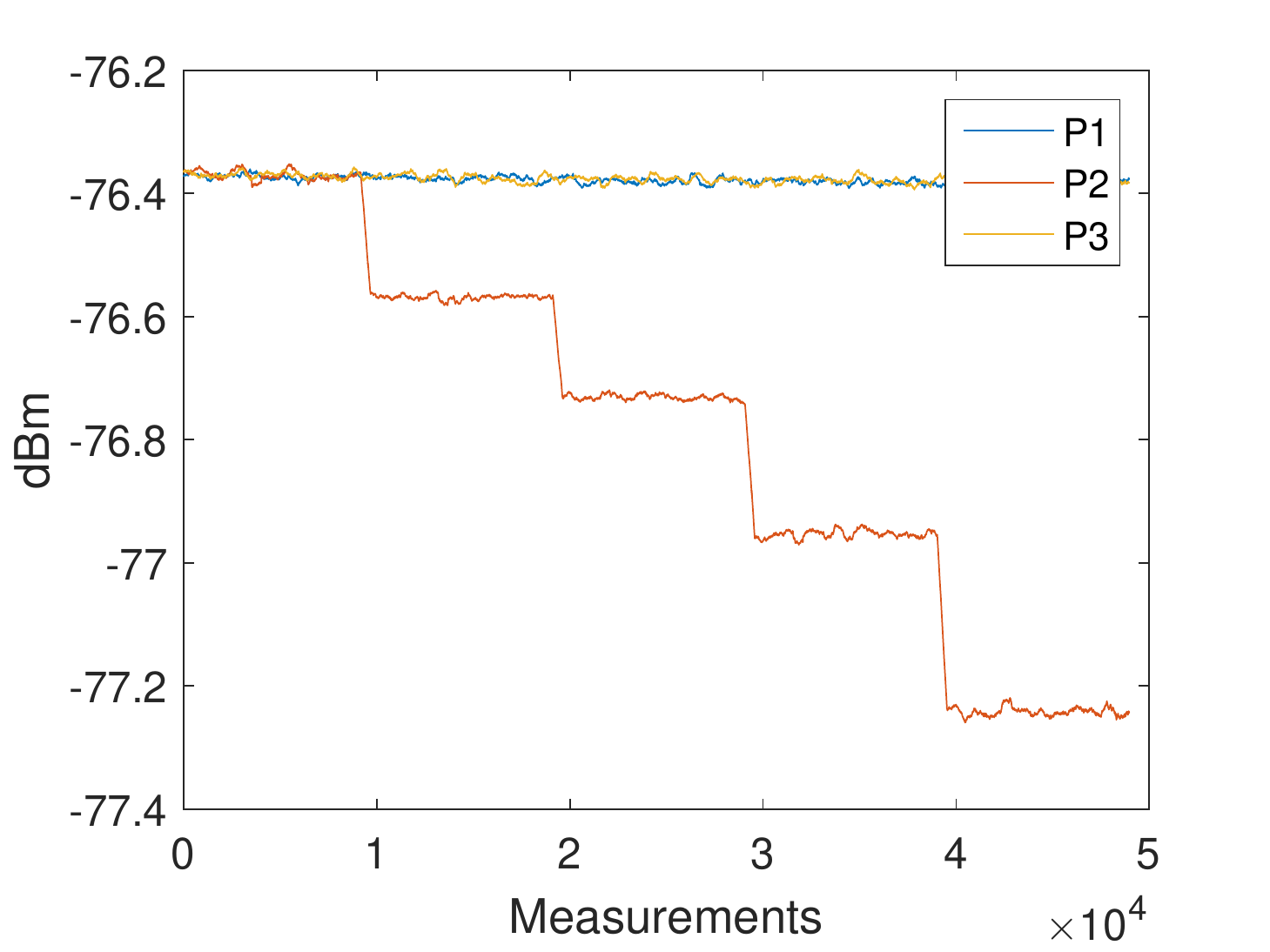}\includegraphics[scale=0.4]{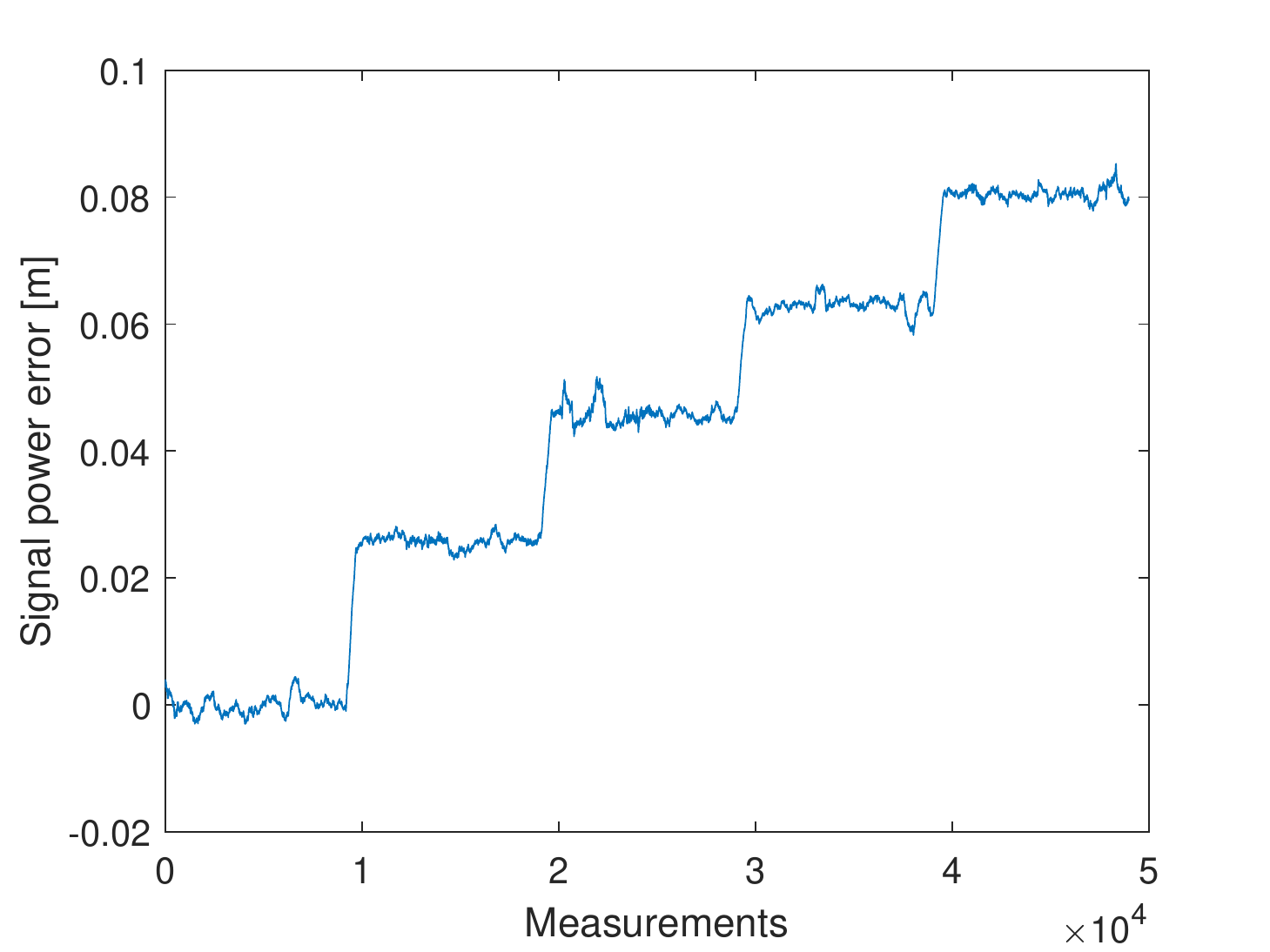}\ 
\par\end{centering}
\caption{Left: Signal strength with cable. Right: Timestamp error with cable\label{fig:Cable-less-then}}
\end{figure}

In the previous section, we discussed an alternative approach to clock
drift correction with three messages (P1, P2, and P3). The following
method is based on this concept, but the TX station changes the signal
strength of the second message (P2). The left side of Figure \ref{fig:Cable-less-then}
shows how the signal strengths of the first and last messages(P1 and
P3) remain constant and only the signal strength of the second signal
(P2) decreases after 1,000 measurements. Every measurement point is
the result of the mean of 2,000 signals. The tests were conducted
with a cable connection of 10 cm and the transmitter decreased the
signal gain with a step size of 3 dB. Figure \ref{fig:Clock-drift-correction}
shows that, after the clock drift correction, the remaining error
of $C_{1,2}^{'}$ (\ref{eq:Correction}) is close to zero. With the
decreasing signal strength of P2, the error of $C_{1,2}$ is increasing;
hence, it is possible to create a dependency between the measured
signal strength and the timestamp error.

\begin{figure}[H]
\begin{centering}
\includegraphics[scale=0.4]{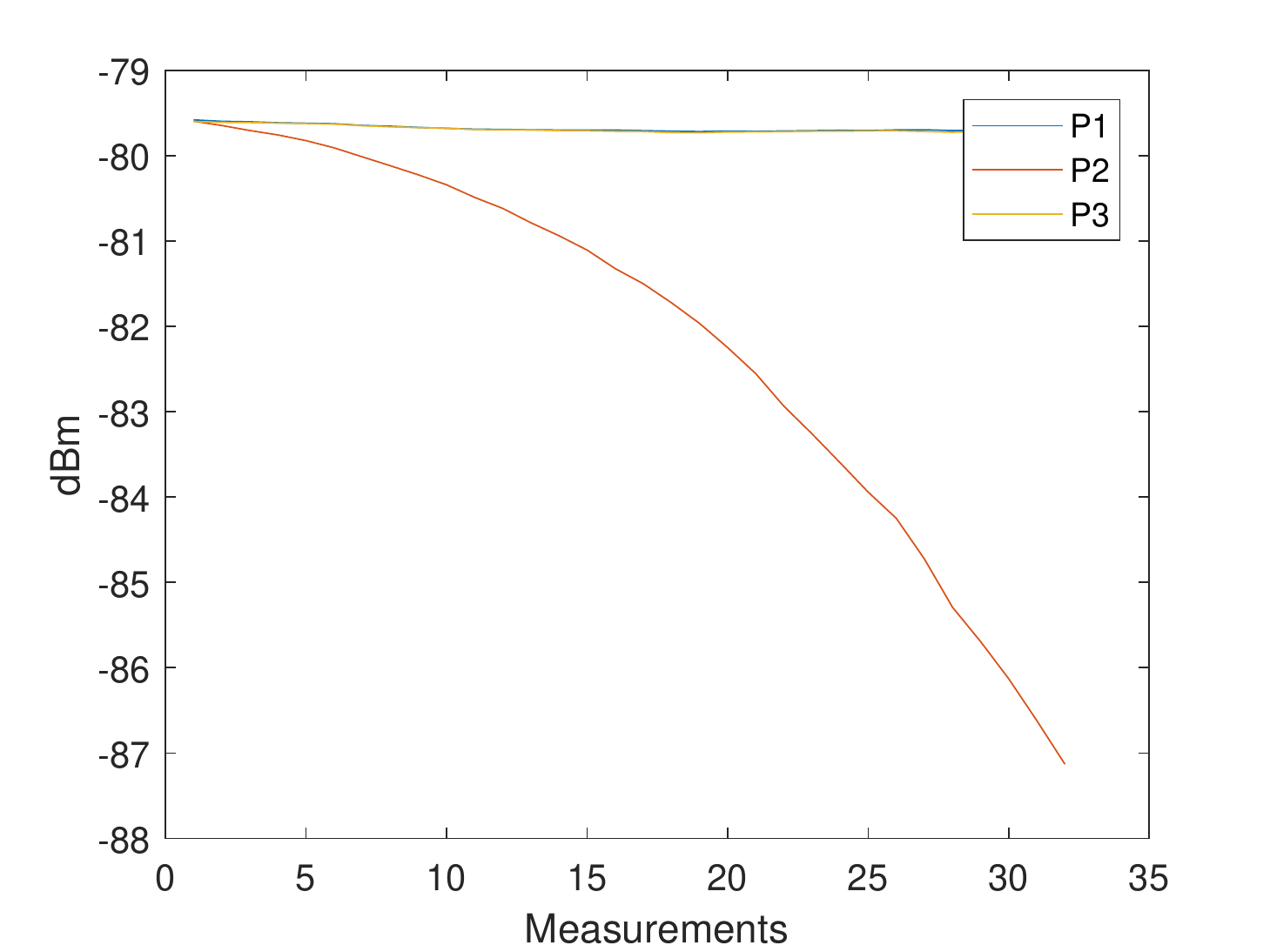}\ \includegraphics[scale=0.4]{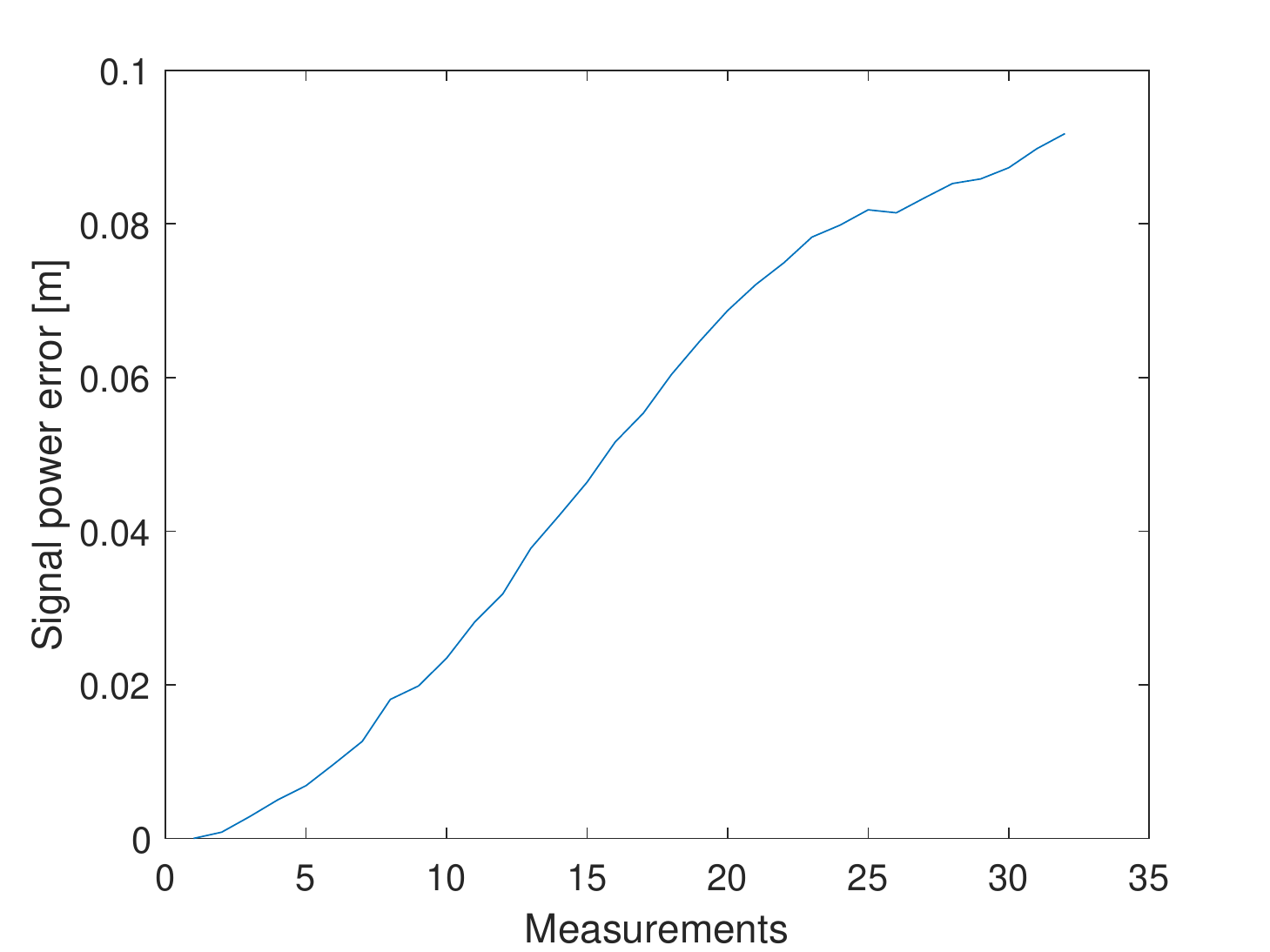}
\par\end{centering}
\caption{Left: Signal strength with antenna. Right: Timestamp error with antenna
\label{fig:Left:-Filtered-signal}}
\end{figure}

In the following test scenario, the power calibration was repeated
with an antenna and a distance of $1.5\,m$ between the RX and TX
stations. The gain step size was reduced to $0.5\,dB$. Figure \ref{fig:Left:-Filtered-signal}
shows the results of the filtered signal power calibration curve.
The main difference between Decawave\textquoteright s curve, as shown
in Figure \ref{fig:Sigmoid}, and our curve is that the zero line
is unknown. This line marks the signal power at which the timestamp
error is zero. The step size of the decreasing transmitting signal
power gain was constant, but the measured decreasing signal power
curve for P2 was nonlinear because the measured signal power did not
equate to the correct signal power for high signal strength, as shown
in Figure \ref{fig:Sigmoid}.

\begin{figure}[H]
\begin{centering}
\includegraphics[scale=0.4]{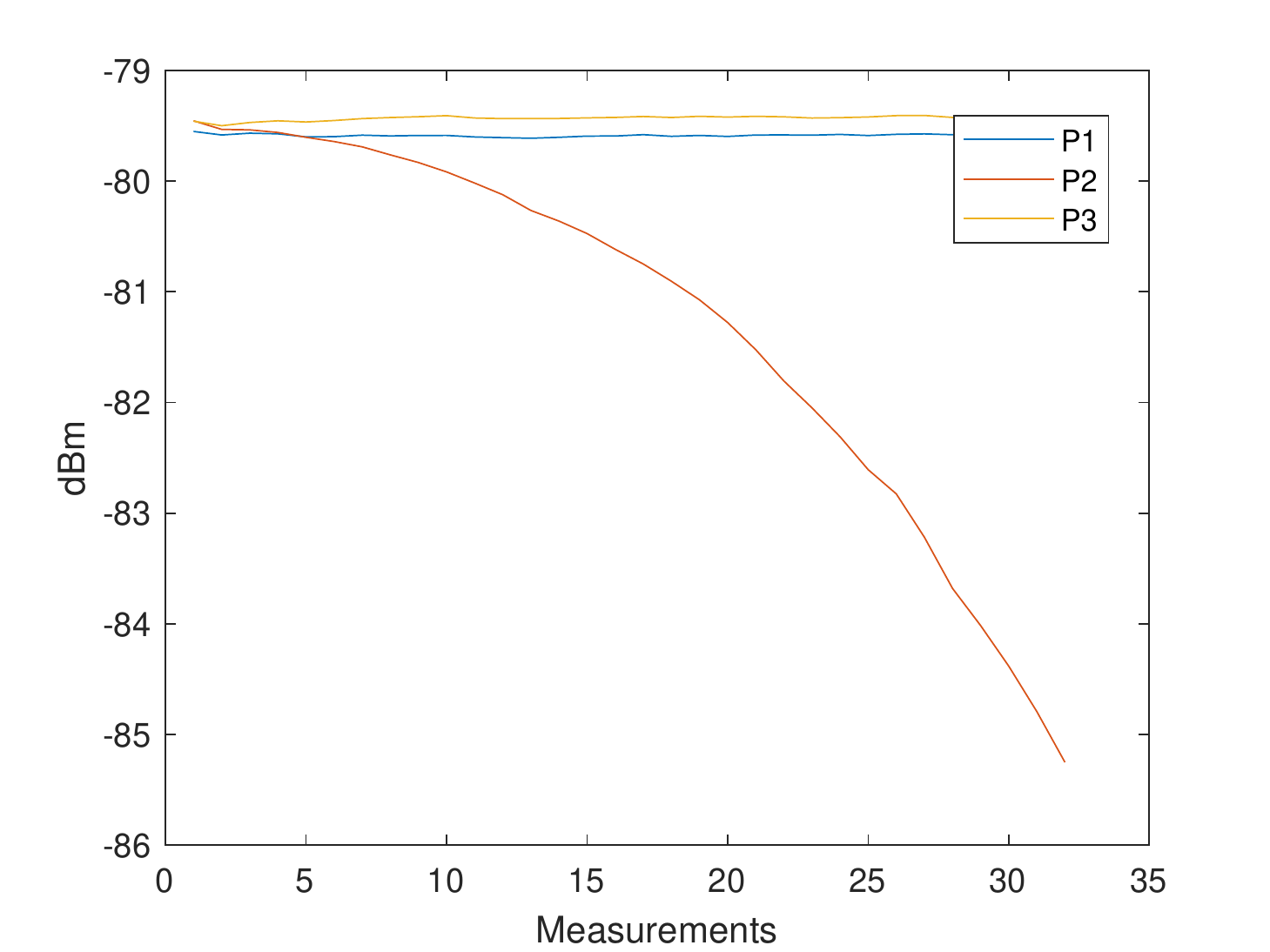}\ 
\par\end{centering}
\caption{Short update time\label{fig:Short-update-time}}
\end{figure}

It is necessary to pay attention to the timing between the messages.
With short delays between the messages, it is possible that they affect
each other. This effect can be seen by the offset between P1 and P3
in Figure \ref{fig:Short-update-time}. In Figure \ref{fig:Left:-Filtered-signal}
a delay of 2 ms has been used between the messages and in Figure \ref{fig:Short-update-time}
a delay of 150 $\mu s$.

\begin{figure}[H]
\begin{centering}
\includegraphics[scale=0.4]{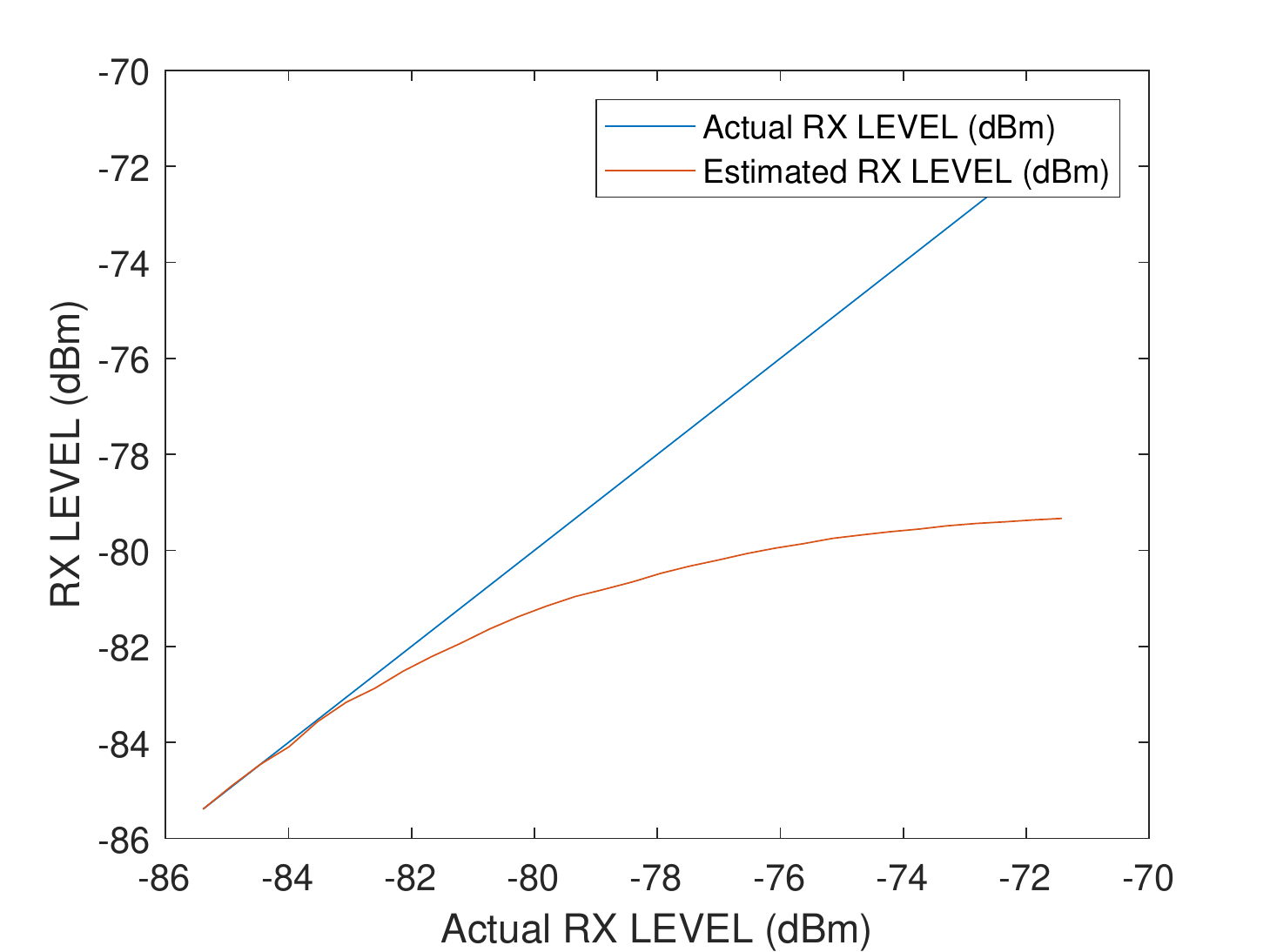}\includegraphics[scale=0.4]{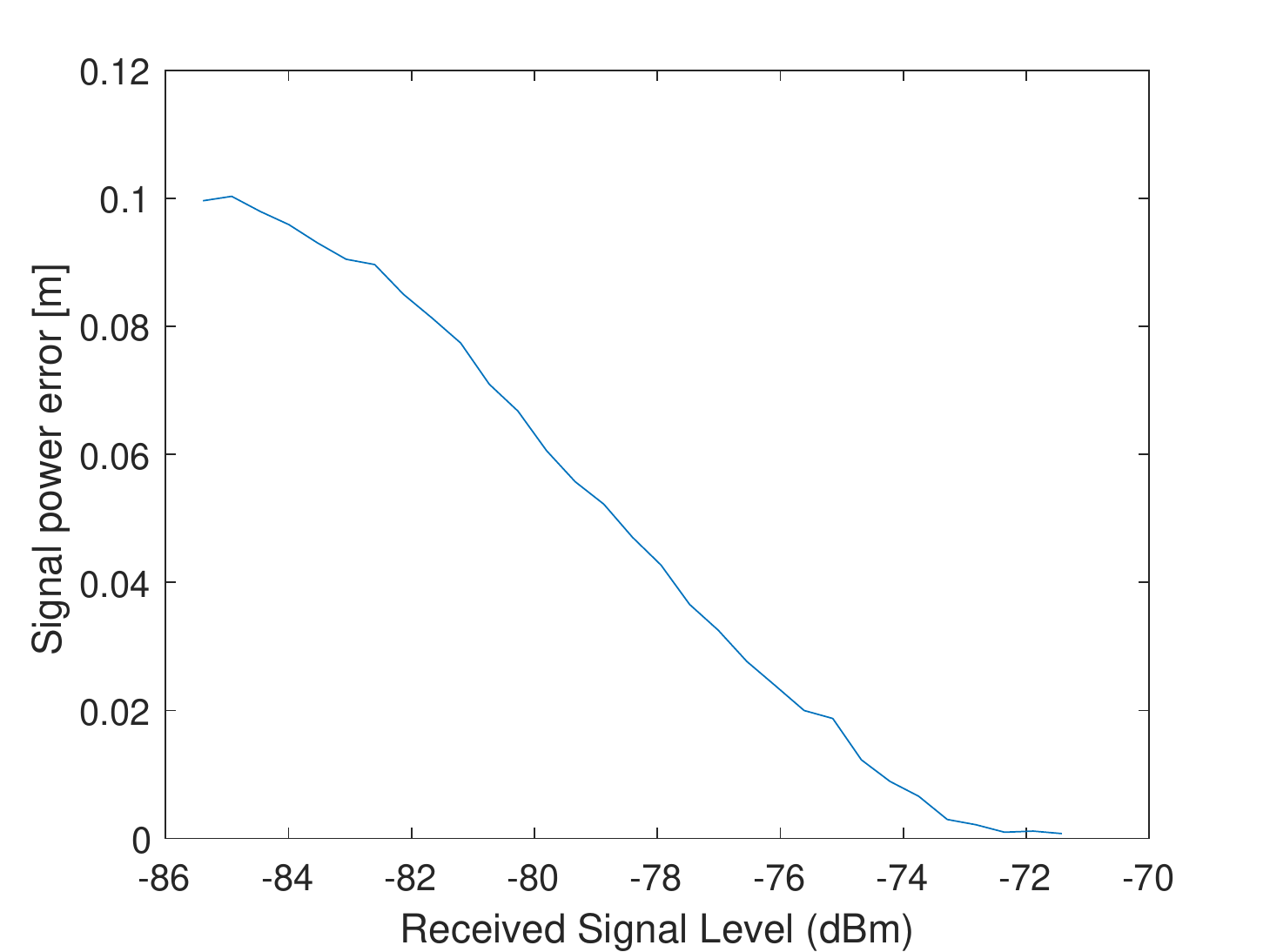}
\par\end{centering}
\caption{Final results of power correction \label{fig:Final-results-of}. Left:
Measured vs. real signal power. Right: Signal power error curve}
\end{figure}

It was previously mentioned that the measured signal strength equals
the correct signal power only for small signal powers. Therefore,
it is possible to use the very first measurements with small signal
strengths to estimate the slope. The left side of Figure \ref{fig:Final-results-of}
shows an estimated line based on the estimated slope. The results
are the same as the curve obtained by Decawave except that no additional
measurement equipment is required and our curves can be obtained individually
for every station. The right side of Figure \ref{fig:Final-results-of}
illustrates the correction curve with respect to the signal power.

\begin{figure}[H]
\begin{centering}
\includegraphics[scale=0.4]{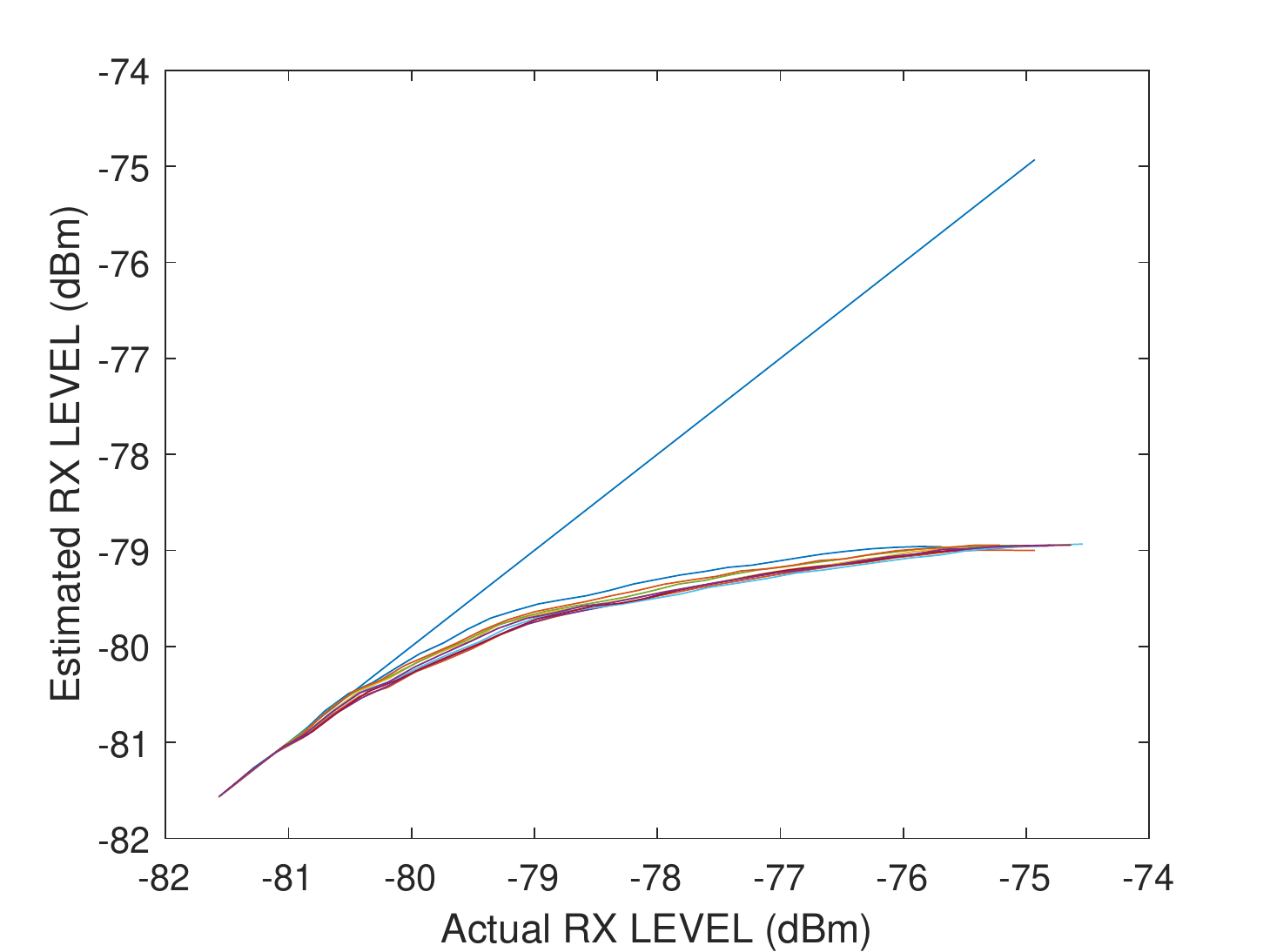}\includegraphics[scale=0.4]{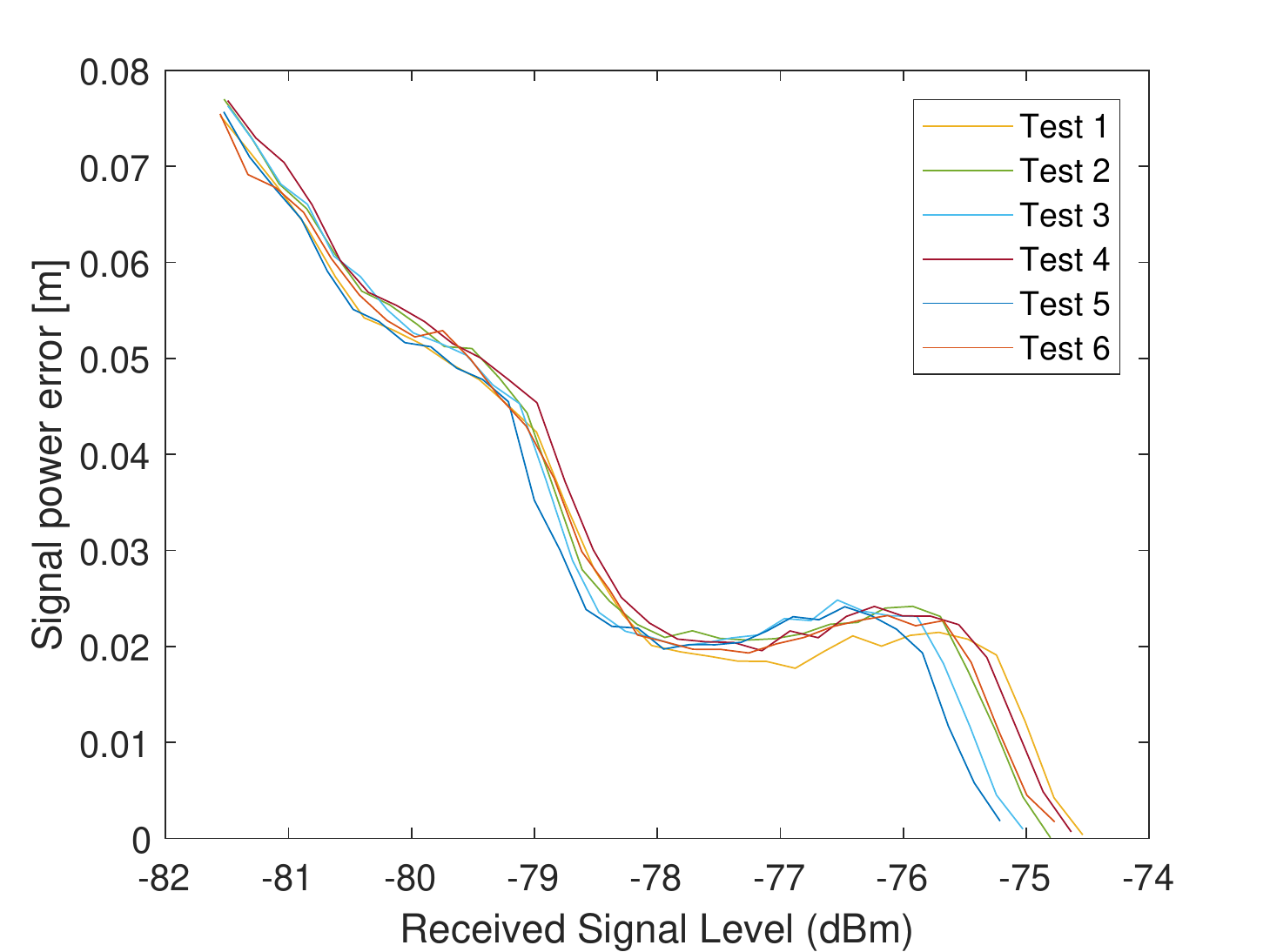}
\par\end{centering}
\caption{Final results of power correction with several restarts\label{fig:Final-results-with}.
Left: Measured vs. real signal power. Right: Signal power error curve}
\end{figure}

Even for the same hardware design, it is possible that the shape of
the correction curve differs. In Figure \ref{fig:Final-results-with},
the final results of the power correction curve are obtained from
another station. The calibration was repeated six times. The shapes
of the curves are deterministic but different from those of the the
station above. Therefore, it makes sense to repeat the calibration
for every individual station. 

\section{Two way ranging \label{sec:Two-way-ranging}}

\begin{figure}[H]
\begin{centering}
\includegraphics[scale=0.45]{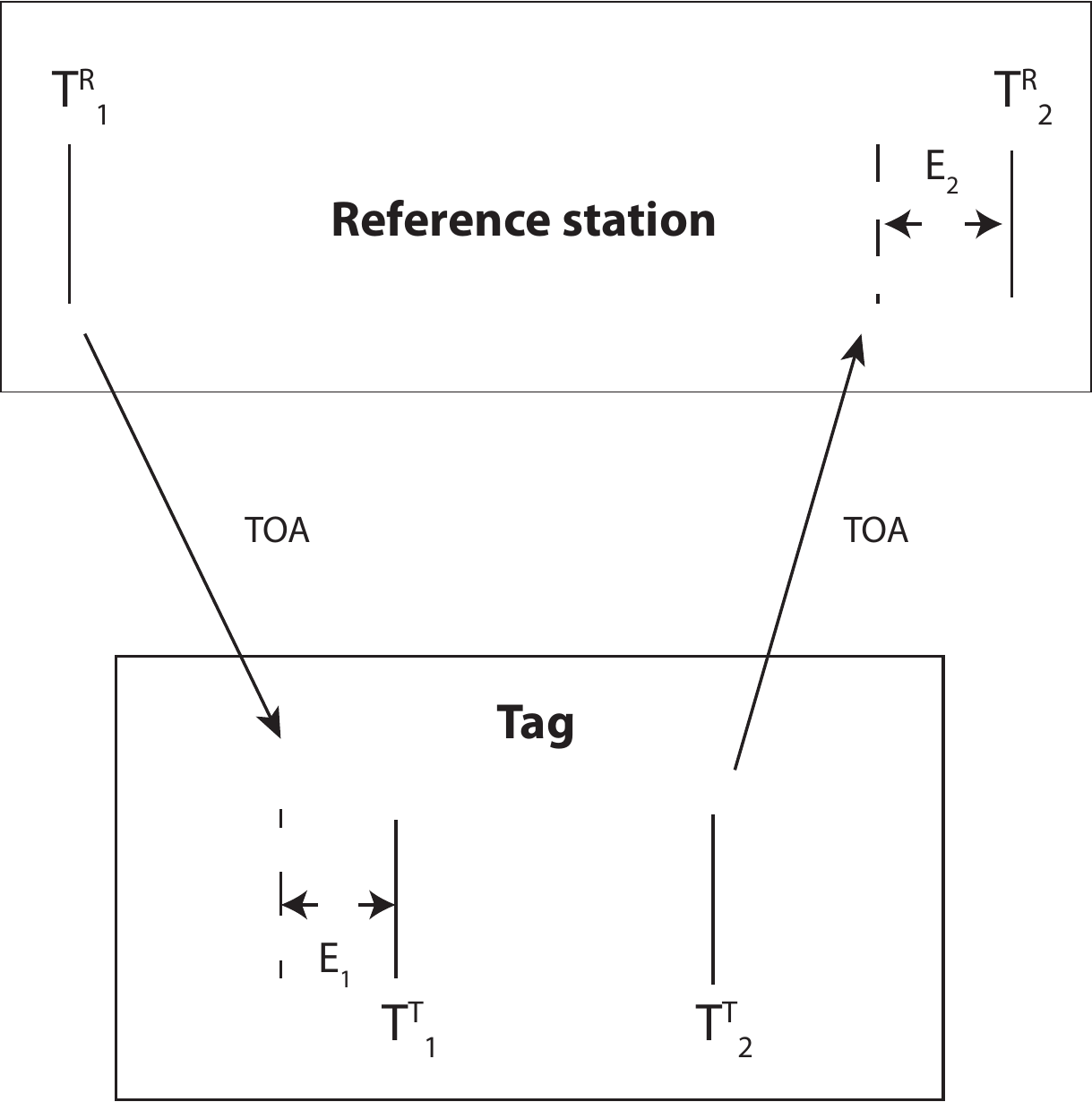}
\par\end{centering}
\caption{Concept for two-way ranging \label{fig:TWR}}
\end{figure}

The following section describes how the presented clock drift and
signal power correction can be used for precise TWR. Figure \ref{fig:TWR}
shows the concept for the TWR. The initial message is sent by the
reference station at $T_{1}^{R}$ and received by the tag. The timestamp
$T_{2}^{T}$ is affected by the signal power and causes an error E1.
After some delay caused by internal processing, the tag sends a response
message at $T_{2}^{T}$. The reference station receives the response
from the tag and saves the timestamp $T_{2}^{R}$, which is affected
by the signal power error E2. In this example, the delay due to the
hardware offset is not considered. 

The time of flight between the reference station and the tag can be
determined by the following formula. It is assumed that the distance
between the two devices does not change between time stamp $T_{2}^{R}$
and $T_{1}^{R}$. 

\begin{equation}
T_{TOA}=\frac{\left(T_{2}^{R}-T_{1}^{R}\right)-\left(T_{2}^{T}-T_{1}^{T}\right)-E_{2}-E_{1}}{2}
\end{equation}

\begin{figure}[H]
\begin{centering}
\includegraphics[scale=0.45]{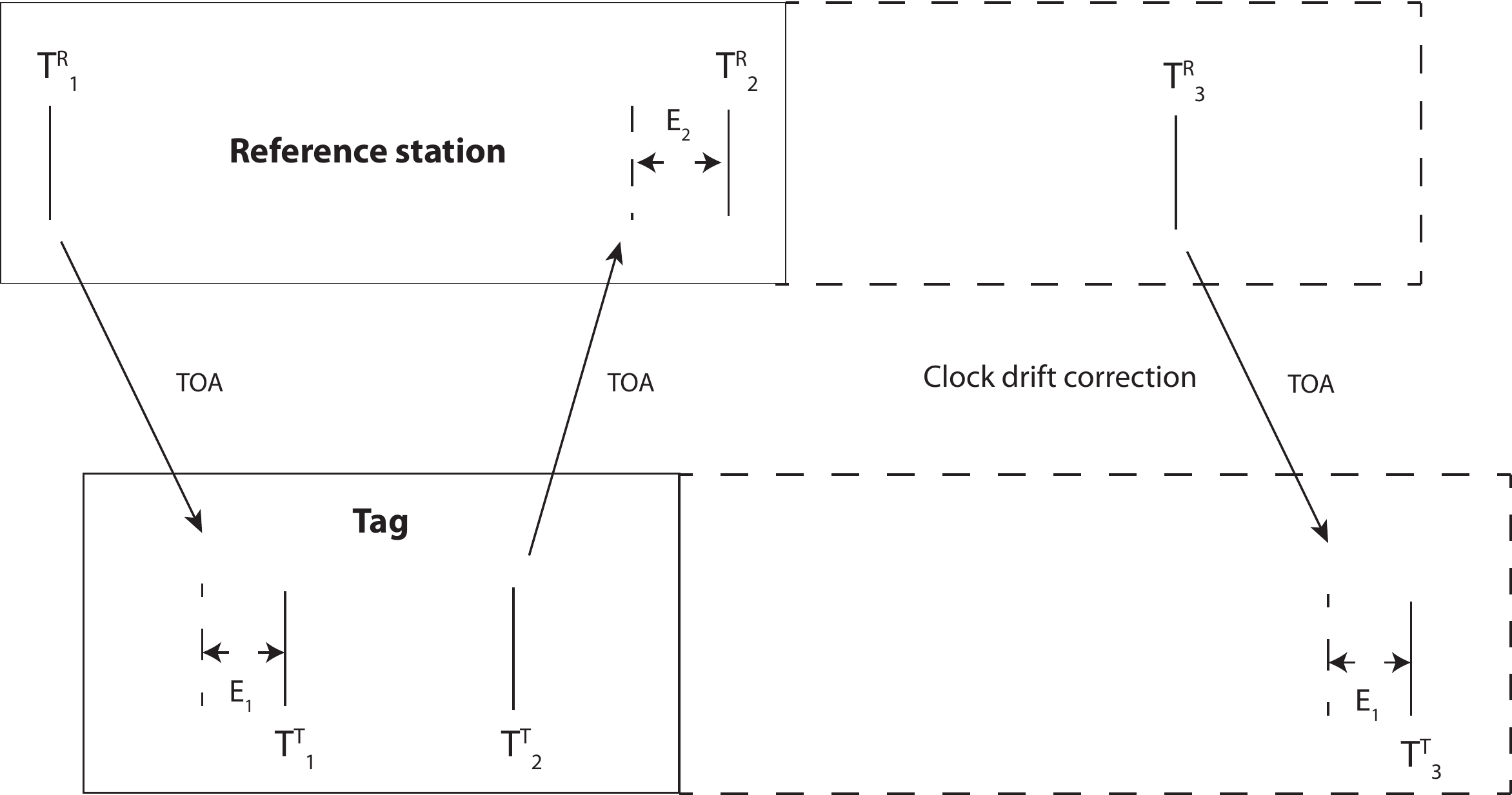}
\par\end{centering}
\caption{TWR clock drift correction \label{fig:TWR-clock-drift}}
\end{figure}

The values E1 and E2 can be obtained from the signal power correction
curve. It should be taken into account that the signal power affects
the tag and reference station differently. The time difference $\Delta T_{1,2}^{R}$
increases with decreasing signal power. The zero lines for both the
signal power and hardware offset are unknown but constant; hence,
both values are represented by the variable Z. In the previous section,
we explained that the clock drift could be corrected by three messages.
Figure \ref{fig:TWR-clock-drift} shows how this principle can be
adapted for TWR. The last message was used to obtain the clock drift
error $C_{1,3}=\Delta T_{1,3}^{R}-\Delta T_{1,3}^{T}$. The signal
power E1 had no effect on the time stamp difference $\Delta T_{1,3}^{T}$.
The final time of the flight equation with the clock drift correction
becomes: 

\begin{equation}
T_{TOA}=0.5\cdotp\left(\Delta T_{1,2}^{R}-\Delta T_{1,2}^{T}-\left(\frac{C_{1,3}^{RT}}{\Delta T_{1,3}^{T}}\cdotp\left(\Delta T_{1,2}^{T}+E1\right)\right)-E_{2}-E_{1}\right)+Z
\end{equation}

\begin{figure}[H]
\begin{centering}
\includegraphics[scale=0.4]{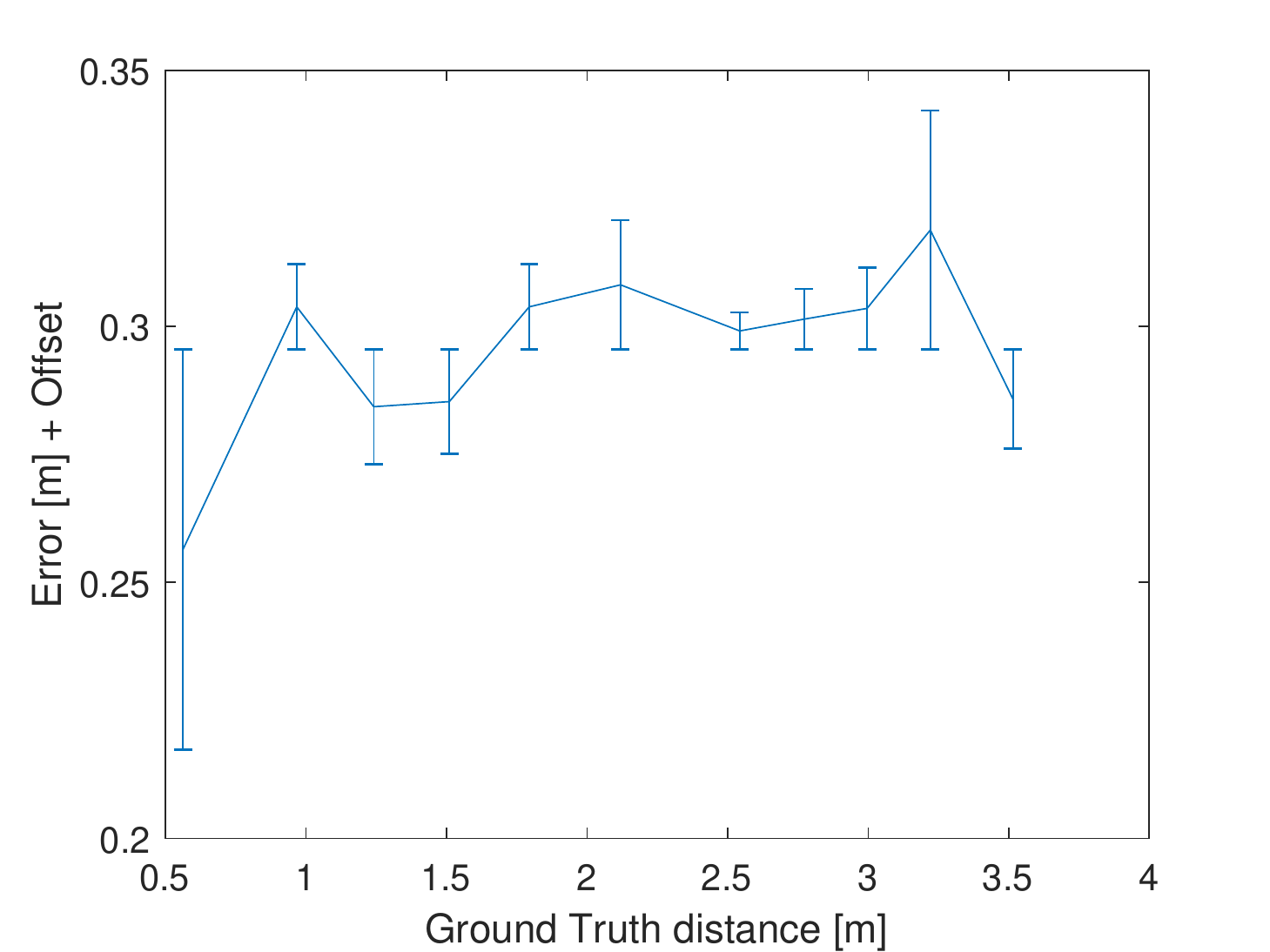}
\par\end{centering}
\caption{TWR test\label{fig:TWR-test}}
\end{figure}

The results of the TWR with signal power and clock drift correction
are illustrated in Figure \ref{fig:TWR-test}. The blue line represents
the difference between laser distance measurements (ground truth)
and distances provided by the TWR. In addition, error bars are used
to illustrate the standard deviation. The 11 distances extend from
$3.515\,m$ to $0.562\,m$. Every point results from the mean of 2,000
measurements. The unkown hardware offset, which cause to the $0.3\,m$
offset, is not relevant in this example. The signal power error depends
on the distance and the clock drift on time. If both effects are corrected
properly the resulting error bars should be as small as possible.
The standard deviation of the error is $0.015\,m$. The small error
bars shows that the signal power and clock drift correction are both
sufficient. The antenna area was $0.0012\,m^{2}$; therefore, it is
not possible to obtain ground truth data with a precision higher than
a few centimeters. 

\section{Conclusion}

This article presents a new method for signal power and clock drift
correction. It was shown that the curves obtained for the signal power
correction could be highly accurate and deterministic, as well as
provide individual results for every station. The signal power correction
procedure can be performed once as a factory calibration. In addition
to the estimation of the signal power correction curve, it was also
possible to obtain the relationship between the measured and real
signal powers. Knowing the relationship allows for better distance
estimations with methods based on the signal strength. In contrast
to the general approach, our clock drift correction is independent
of the signal power and promises results with centimeter accuracy.
The last part of the article explained how the signal power and clock
drift correction are fused together to provide highly accurate TWR.

\bibliographystyle{unsrt}
\bibliography{uwb_power}

\end{document}